\documentclass[twocolumn,aps,showpacs]{revtex4}

\usepackage{amsmath}
\usepackage{amssymb}
\usepackage{graphicx}
\usepackage{dcolumn}

\begin{document}

\title{Theory of magnetothermoelectric phenomena in high-mobility two-dimensional 
electron systems under microwave irradiation}

\author{O. E. Raichev}
\affiliation{Institute of Semiconductor Physics, National Academy of Sciences of Ukraine, 
Prospekt Nauki 41, 03028, Kyiv, Ukraine}
\date{\today}

\begin{abstract}
The response of two-dimensional electron gas to temperature gradient 
in perpendicular magnetic field under steady-state microwave irradiation 
is studied theoretically. The electric currents induced by temperature 
gradient and the thermopower coefficients are calculated taking into account 
both diffusive and phonon-drag mechanisms. The modification of thermopower 
by microwaves takes place because of Landau quantization of electron energy 
spectrum and is governed by the microscopic mechanisms which are similar to 
those responsible for microwave-induced oscillations of electrical resistivity. 
The magnetic-field dependence of microwave-induced corrections to phonon-drag 
thermopower is determined by mixing of phonon resonance frequencies with 
radiation frequency, which leads to interference oscillations. The transverse 
thermopower is modified by microwave irradiation much stronger than the 
longitudinal one. Apart from showing prominent microwave-induced oscillations 
as a function of magnetic field, the transverse thermopower appears to be 
highly sensitive to the direction of linear polarization of microwave 
radiation.   
\end{abstract}


\maketitle

\section{Introduction}

Electron transport in two-dimensional (2D) electron systems placed in a perpendicular magnetic 
field remains one of the most important subjects in the condensed matter physics. Recently, 
it was established that a variety of interesting transport phenomena takes place$^1$ in the 
region of moderately strong magnetic field, where the Shubnikov-de Haas oscillations of the 
electrical resistivity are suppressed because of thermal smearing of the Fermi level. 
In particular, there are several kinds of magnetoresistance oscillations$^1$ observed in 
high-mobility 2D systems such as GaAs quantum wells, strained Ge quantum wells, and electrons 
on liquid helium surface. Among these phenomena, the microwave-induced resistance oscillations 
(MIRO) appearing under microwave (MW) irradiation of 2D electron gas$^{2-5}$ are studied 
most extensively. Their origin is briefly described as follows. In the presence of the MW 
excitation, when absorption and emission of radiation quanta by electron system take 
place, both the distribution function and scattering probabilities of electrons are modified. 
These modifications correlate with the oscillating density of electron states owing to Landau 
quantization in the magnetic field $B$, thus leading to corresponding oscillating contributions 
to resistivity determined by the ratio of the radiation frequency $\omega$ to the cyclotron 
frequency $\omega_{c}=|e|B/mc$. The period and phase of MIRO, as well as the temperature and 
power dependence of their amplitudes are in agreement with this physical picture supported by 
a detailed consideration of microscopic mechanisms of MIRO in the past years.$^{6-11}$ 
According to both experiment and theory, the MW irradiation strongly affects the longitudinal 
(dissipative) resistivity and has a much weaker effect on the transverse (Hall) resistivity. 
More recent experiments uncover the existence of small corrections, sensitive to the 
direction of electric field of microwaves (MW polarization), to both longitudinal and 
Hall resistivities,$^{12,13}$ also in general agreement with the theory.  

Apart from its influence on electrical resistivity, the MW excitation is expected to 
modify other transport coefficients of 2D electrons, for the same reasons as explained 
above. The thermoelectric coefficients are of special interest in this connection. 
The study of thermoelectric phenomena in magnetic fields has a long history, and the 
fundamentals of this topic, with applications to bulk conductors, are reviewed in 
Ref. 14. The electrical response to temperature gradient $\nabla T$ is 
described by the longitudinal (Seebeck) and transverse (Nernst-Ettingshausen) components 
of thermoelectric power (briefly, thermopower). These coefficients are determined by two 
mechanisms: the diffusive one, when electrons are directly driven by the diffusion force 
due to temperature gradient in electron gas, and phonon drag one, when electrons are driven 
by a frictional force between them and phonons propagating along the temperature gradient. 
Contribution of both these mechanisms in magnetothermopower of 2D electron systems has been 
studied in a number of theoretical and experimental works$^{15-21}$ (see also review 
paper Ref. 22 and references therein). The quantum effects are commonly observed 
in strong magnetic fields, where Shubnikov-de Haas oscillations of thermopower 
coefficients take place.$^{22}$ In high-mobility GaAs quantum wells, the 
phonon-drag thermopower shows another kind of quantum oscillations, related to resonant 
phonon-assisted scattering of electrons between Landau levels.$^{20}$ This occurs in the 
region of moderately strong magnetic fields, below the onset of the Shubnikov-de Haas 
oscillations, which is favorable for observation of MW-induced quantum effects. 

There are two main ways in which the MW irradiation can influence the thermopower. 
First, this irradiation leads to non-equilibrium electron distribution 
that has non-trivial dependence not only on electron energy but also on temperature 
of electron gas. Both the diffusive and phonon-drag contributions to thermoelectric 
coefficients should be sensitive to such changes. Next, the MW irradiation in the 
presence of magnetic field considerably influences electron-phonon scattering. 
This causes an effect on electrical resistivity$^{23}$ under conditions when 
probability of electron-phonon scattering is comparable to that of elastic 
scattering by impurities. At the temperatures below 4.2 K these conditions are 
realized only in very pure 2D systems. In contrast, the effect of microwaves on 
electron-phonon scattering is always important for thermoelectric properties, 
since the phonon drag mechanism determined by this scattering gives a very 
significant,$^{15}$ if not a major, contribution to thermopower of 2D electrons. 

The above consideration also suggests that in spite of the same microscopic mechanisms 
involved in both cases, the effect of microwaves on magnetothermoelectric coefficients 
of 2D electrons should be different from their effect on magnetoresistance. The 
classical Mott relation between the diffusive current responses to temperature 
gradient and to electric field is not expected to be valid under MW excitation, 
even for the moderately strong magnetic fields. Moreover, one may presume that both 
longitudinal and transverse components of thermopower oscillate with magnetic field 
in the way different from MIRO, and their dependence on MW polarization is also
different. Therefore, there is enough motivation for a theoretical study of the 
influence of MW irradiation on thermopower of 2D electron systems in perpendicular 
magnetic field. The present paper is devoted to this previously unexplored problem. 
   
In the linear response regime considered in the following, the current density 
${\bf j}$ is given by the general expression
\begin{equation}
{\bf j}= \hat{\sigma} {\bf E} - \hat{\beta} \nabla T, 
%1
\end{equation}
where ${\bf E}$ is the electric field in the plane $(x,y)$ of the 2D electron system. It is 
assumed that the 2D system is macroscopically homogeneous so that the chemical potential 
$\mu$ does not depend on 2D coordinate ${\bf r}$. Under condition when no conduction 
currents flow in the system, one gets ${\bf E} = \hat{\alpha} \nabla T$. The 
thermopower tensor $\hat{\alpha}$ describes the voltage drop as a result of temperature 
gradient. It is given by $\hat{\alpha}= \hat{\rho} \hat{\beta}$, where the resistivity 
tensor $\hat{\rho}$ is the matrix inverse of the conductivity tensor $\hat{\sigma}$. 
The theoretical approach presented below is based upon calculation of thermoelectric tensor 
$\hat{\beta}$ in the presence of ac field of microwaves by using the method of quantum Boltzmann 
equation$^{1,8,10,23}$ established in the previous calculations of the conductivity 
tensor $\hat{\sigma}$. As both $\hat{\beta}$ and $\hat{\sigma}$ are known, the 
thermopower is found straightforwardly. The results are presented for the case of 
moderately strong magnetic field, when the Shubnikov-de Haas oscillations are still 
suppressed, but quantum oscillations due to Landau quantization exist in high-mobility 
2D systems. Such oscillations are caused by inelastic scattering of electrons between 
Landau levels as a result of electron interaction with acoustic phonons of a
resonant frequency $\omega_{ph}$ (magnetophonon effect$^{20,24-29}$) and with 
microwaves of frequency $\omega$. These two kinds of inelastic processes actually 
interfere, leading to combined frequencies $\omega_{ph} \pm \omega$ whose ratio 
to $\omega_c$ determines the periodicity of the quantum magnetooscillations.$^{23}$ 
As shown below, such oscillations exist in both longitudinal and transverse thermopower 
caused by the phonon drag, while the diffusion part of the thermopower follows the 
MIRO periodicity determined by the single frequency $\omega$. The phonon-drag 
part of MW-induced contribution to transverse thermopower is found to be 
comparable with that of longitudinal thermopower. Since the transverse thermopower
is much smaller than the longitudinal one in classically strong magnetic fields, 
it is dramatically affected by MW irradiation, demonstrating giant microwave-induced 
oscillations and a high sensitivity to the direction of MW polarization.      

The paper is organized as follows. Section II describes the main formalism including 
description of ac electric field generated by incident electromagnetic radiation, 
electric current in the presence of temperature gradient, and kinetic equation for 2D 
electrons with collision integrals for electron-impurity and electron-phonon 
interactions. In Section III the kinetic equation is solved and the tensor 
$\hat{\beta}$ is presented and discussed both for the equilibrium case and under MW 
irradiation. Section IV contains expressions for longitudinal and transverse components 
of thermopower tensor $\hat{\alpha}$, their discussion, and presentation of the results 
of numerical calculations of these components as functions of magnetic field and 
polarization angle. More discussion and concluding remarks are given in the last section.
 
\section{General formalism}

Throughout the paper, one uses the system of units where Planck's constant $\hbar$ 
and Boltzmann constant $k_B$ are both set to unity. The electron spectrum is assumed 
to be isotropic and parabolic, with effective mass $m$. The Zeeman splitting of 
electron states is neglected.

Consider a monochromatic electromagnetic wave normally incident on the surface 
containing a 2D layer (the direction of incidence coincides with the direction 
of the magnetic field, along the $z$ axis). The electric field of this wave near 
the layer is written, in the general form, as  
\begin{eqnarray}
{\bf E}^{(i)}_t=E^{(i)}_{\omega} {\rm Re} [ {\bf e} e^{-i \omega t}]
\nonumber \\
=\frac{E^{(i)}_{\omega}}{\sqrt{2}} 
{\rm Re} \left\{ \left[ {\rm e}_-\left(\begin{array}{c} 1 \\ i \end{array} \right) 
+ {\rm e}_+ 
\left(\begin{array}{c}1 \\ -i \end{array} \right) \right] e^{-i \omega t} \right\},
%2
\end{eqnarray}
where ${\bf e}$ is the polarization vector. The second part of this equation represents
the wave as a sum of two circularly polarized waves, ${\rm e}_{\pm}= ({\rm e}_x \pm i 
{\rm e}_y)/\sqrt{2}= \kappa_{\pm} e^{\pm i \chi}$, $\chi$ is the angle between the 
main axis of polarization of ${\bf E}^{(i)}_t$ and the $x$ axis, and $\kappa_{\pm}$ 
are real numbers characterizing ellipticity of the incident wave (they are normalized 
according to $\kappa^2_{+} + \kappa^2_{-}=1$). A circular polarization means that 
either $\kappa_{+}$ or $\kappa_{-}$ is equal to zero. In the case of linear polarization, $\kappa_{+}=\kappa_{-}=1/\sqrt{2}$ so that ${\rm e}_{\pm}=e^{\pm i \chi}/\sqrt{2}$. 
The screening of electromagnetic wave due to the presence of free carriers in the 2D layer 
changes the polarization angle and ellipticity,$^{30,31}$ so the electric field in the layer, 
${\bf E}_t$, differs from ${\bf E}^{(i)}_t$ and has the following form:
\begin{eqnarray}
{\bf E}_t=\frac{E_{\omega}}{\sqrt{2}} 
{\rm Re} \left\{ \left[ (\omega - \omega_c) s_- \left(\begin{array}{c} 1 \\ i \end{array} \right) 
\right. \right. \nonumber \\ 
\left. \left. + (\omega + \omega_c) s_+ \left(\begin{array}{c}1 \\ -i \end{array} \right)  \right]
e^{-i \omega t} \right\},
%3
\end{eqnarray} 
where 
\begin{equation}
s_{\pm}= \frac{{\rm e}_{\pm}}{\omega \pm \omega_c + i \omega_p}.
%4
\end{equation}
Here $\omega_p$ is the radiative decay rate that determines the cyclotron line broadening 
because of electrodynamic screening effect. It is given by $\omega_p=2 \pi e^2 n_s/ m c 
\sqrt{\epsilon^*}$, where $n_s$ is the electron density, $\sqrt{\epsilon^*}=(1+ 
\sqrt{\epsilon})/2$, and $\epsilon$ is the dielectric permittivity of the sample 
material. Next, $E_{\omega}=E^{(i)}_{\omega}/\sqrt{\epsilon^*}$. In Eqs. (3) and (4), 
it is assumed that transport relaxation rate, $\nu_{tr}$, which determines electron 
mobility, is much smaller than either $|\omega \pm \omega_c|$ or $\omega_p$. The 
relation $\nu_{tr} \ll \omega_p$ is a very good approximation for high-mobility 
samples with typical electron density $n_s > 10^{11}$ cm$^{-2}$.

Apart from the ac field ${\bf E}_t$, the electron system is driven by a weak static (dc) field
${\bf E}$. To take into account the influence of both these fields on 2D electrons, it is very 
convenient to use a transition to the moving coordinate frame (see Ref. 10 and references therein). 
Then, the quantum kinetic equation for electrons derived by using Keldysh formalism for non-equilibrium 
electron systems (see details in Refs. 10 and 23) contains the effect of external fields 
only in the collision integral. The radiation power is assumed to be weak enough to neglect the 
influence of microwaves on the energy spectrum of electrons: the spectrum remains isotropic and 
the density of states is not affected by the radiation. Further, the magnetic field is assumed 
to be weak enough so there is a large number of Landau levels under the Fermi energy. The 
kinetic equation written for the electron distribution function $f_{\varepsilon \varphi}$ 
averaged over the period $2\pi/\omega$ takes the form 
\begin{equation}
\frac{{\bf p}_{\varepsilon \varphi}}{m} \cdot \nabla f_{\varepsilon \varphi}
+ \omega_c \frac{\partial f_{\varepsilon \varphi}}{ \partial \varphi}  
=J_{\varepsilon \varphi},~~~J_{\varepsilon \varphi} = J^{im}_{\varepsilon \varphi}+ 
J^{ph}_{\varepsilon \varphi},
%5
\end{equation}
where $\varepsilon$ is the electron energy, ${\bf p}_{\varepsilon \varphi}=p_{\varepsilon
}(\cos \varphi, \sin \varphi)$ with $p_{\varepsilon}=\sqrt{2m \varepsilon}$ is the electron 
momentum in the 2D layer plane, and $\varphi$ is the angle of this momentum. Since the dependence 
of all quantities on the spatial coordinate ${\bf r}$ is considered as a parametric one, the 
coordinate index at the distribution function and collision integrals is omitted. 
The density of electric current is given by the expression
\begin{eqnarray}
{\bf j} =\frac{e}{\pi} \int d \varepsilon D_{\varepsilon} \int_0^{2 \pi} \frac{d \varphi}{2 \pi} 
{\bf p}_{\varepsilon \varphi} f_{\varepsilon \varphi} - \sigma_{\bot} \hat{\epsilon} {\bf E} 
-c \hat{\epsilon} \nabla M_z,
%6
\end{eqnarray} 
where $\sigma_{\bot}=e^2 n_s /m \omega_c= |e|n_s c/B$ is the classical Hall conductivity 
and $D_{\varepsilon}$ is the density of electron states expressed in the units $m/\pi$. 
Next, $\hat{\epsilon}=\left( \begin{array}{cc} 0 & 1 \\ -1 & 0 \end{array} \right)$ 
is the antisymmetric unit matrix in the space of 2D Cartesian indices. The last term in 
the expression (6) is given by the spatial gradient of magnetic moment ${\bf M}$ 
of electrons per unit square. This moment arises because of diamagnetic currents 
circulating in the electron system. Actually, the last term in Eq. (6) does not 
contribute to the {\em total} current across any finite sample. However, the necessity 
of taking into account this term (its bulk analogue is $-c [\nabla \times {\bf M}]$) 
in the expression for the {\em local} current density has been emphasized in studies of 
magnetothermoelectric phenomena a long time ago.$^{14,32}$ Being expressed through the 
distribution function, the magnetic moment comprises two terms:
\begin{equation}
M_z = -\frac{m}{\pi B} \int d \varepsilon [D_{\varepsilon} \varepsilon - \Pi_{\varepsilon}]
f_{\varepsilon},
%7
\end{equation} 
where $f_{\varepsilon}$ is the isotropic (averaged over $\varphi$) part of electron distribution 
function, and $\Pi_{\varepsilon}= \int_{-\infty}^{\varepsilon} d \varepsilon' D_{\varepsilon'}$ 
is the antiderivative of $D_{\varepsilon}$. In the ideal 2D electron system, the first and 
the second terms in $M_z$ correspond to magnetization due to bulk and edge currents, 
respectively.$^{33}$ In the case of the equilibrium Fermi distribution function 
$f_{\varepsilon}= \{\exp[(\varepsilon-\mu)/T]+1 \}^{-1}$, it is easy to transform Eq. (7) to 
a well-known thermodynamic expression $M_z=-\partial \Omega/\partial B$, where $\Omega=-(T m/\pi) 
\int d \varepsilon D_{\varepsilon} \ln \left\{1+\exp[(\mu -\varepsilon)/T] \right\}$ is the 
thermodynamic potential per unit area. 

In the absence of any collisions, $J_{\varepsilon \varphi}=0$, the local current is 
non-dissipative, ${\bf j}={\bf j}^{(n)}$, where
\begin{equation}
{\bf j}^{(n)}= -\frac{c m}{\pi B} \int d \varepsilon \Pi_{\varepsilon}
\hat{\epsilon} \nabla f_{\varepsilon} - \sigma_{\bot} \hat{\epsilon} {\bf E}.
%8
\end{equation}
If coordinate dependence of $f_{\varepsilon}$ exists solely due to temperature gradient, 
one has $\nabla f_{\varepsilon}= (\partial f_{\varepsilon}/\partial T) \nabla T$.
The integral term in Eq. (8) is reduced to non-dissipative thermoelectric current 
$-\hat{\beta} \nabla T$ flowing perpendicular to $\nabla T$, with $\hat{\beta}=
(c m/\pi B) \int d \varepsilon \Pi_{\varepsilon} (\partial f_{\varepsilon}/\partial T) 
\hat{\epsilon}$. If chemical potential $\mu$ entering $f_{\varepsilon}$ also depends on 
coordinate, the integral in Eq. (8) produces an additional term proportional to 
$\nabla \mu$. This term, with the aid of the identity $\partial f_{\varepsilon}/\partial 
\mu = -\partial f_{\varepsilon}/\partial \varepsilon$, can be combined with the last term 
of Eq. (8), leading to the form $\sigma_{\bot} \hat{\epsilon} \nabla \zeta$, where 
$\zeta=\Phi+\mu/e$ is the electrochemical potential and $\Phi$ is the electrostatic 
potential determining the electric field ${\bf E}= - \nabla \Phi$. The electric 
field or, in general, the gradient of electrochemical potential induced as a result of 
a temperature gradient is derived from the expression ${\bf j}^{(n)}=0$. This leads to 
diagonal thermopower tensor $\hat{\alpha}= \hat{1} \alpha$, where $\hat{1}$ is the unit 
$2 \times 2$ matrix and
\begin{equation}
\alpha = \frac{m}{\pi e n_s} \int d \varepsilon \Pi_{\varepsilon} 
\frac{\partial f_{\varepsilon}}{\partial T}.
%9
\end{equation}
Substituting the equilibrium distribution function into Eq. (9), one gets a well-known result
\begin{equation}
\alpha = -\frac{S}{|e| n_s}, ~~S=-\frac{\partial \Omega}{\partial T},
%10
\end{equation}   
where $S$ is the entropy of 2D electron gas per unit area. For strongly degenerate 
electron gas, $\mu=\varepsilon_F$, one has $S=(\pi^2/3)n_s T/\varepsilon_F$. 
 
The collision integrals $J^{im}_{\varepsilon \varphi}$ and $J^{ph}_{\varepsilon \varphi}$ standing
in Eq. (5) describe, respectively, electron-impurity and electron-phonon scattering:$^{23}$ 
\begin{eqnarray}
J^{im}_{\varepsilon \varphi}=\int_0^{2 \pi} \frac{d \varphi'}{2 \pi}
\sum_{n=-\infty}^{\infty} \nu(|{\bf q}_{\varepsilon n}|) 
[J_n (|{\bf R}_{\omega} \cdot {\bf q}_{\varepsilon n}|)]^2 \nonumber \\ 
\times D_{\varepsilon+n \omega+\gamma_n} [f_{\varepsilon+n \omega+\gamma_n \varphi'} - f_{\varepsilon \varphi}],
%11
\end{eqnarray} 
\begin{eqnarray}
J^{ph}_{\varepsilon \varphi}=\int_0^{2 \pi} \frac{d \varphi'}{2 \pi} 
\sum_{\lambda} \int_{-\infty}^{\infty} \frac{d q_z}{2 \pi}
m  \sum_{n=-\infty}^{\infty} \nonumber \\
\times \left\{ M_{\lambda {\bf Q}^-} [J_n (|{\bf R}_{\omega} \cdot {\bf q}^-_{\varepsilon n}|)]^2 
[(N_{\lambda {\bf Q}^-}+f_{\varepsilon \varphi}) 
 \right. \nonumber \\
\left. \times f_{\varepsilon - \omega_{\lambda {\bf Q}^-} + n \omega +\gamma^-_n \varphi'} 
- (N_{\lambda {\bf Q}^-}+1) f_{\varepsilon \varphi} ] \right. \nonumber \\
\times \left.
D_{\varepsilon - \omega_{\lambda {\bf Q}^-} + n \omega +\gamma^-_n} 
 + M_{\lambda {\bf Q}^+} [J_n (|{\bf R}_{\omega} \cdot {\bf q}^+_{\varepsilon n}|)]^2 
\right. \nonumber \\
\left.  
\times [(N_{\lambda -{\bf Q}^+}+ 1 -f_{\varepsilon \varphi})  
f_{\varepsilon + \omega_{\lambda {\bf Q}^+} + n \omega +\gamma^+_n \varphi'} 
\right. \nonumber \\
\left.  
- N_{\lambda -{\bf Q}^+} f_{\varepsilon \varphi} ] 
D_{\varepsilon + \omega_{\lambda {\bf Q}^+} + n \omega +\gamma^+_n}\right\},
%12
\end{eqnarray} 
where $J_n$ is the Bessel function, $\nu(q)=m w(q)$ is the isotropic elastic scattering 
rate expressed through the Fourier transform $w(q)$ of the correlation function of 
random potential of impurities,
${\bf q}_{\varepsilon n}={\bf p}_{\varepsilon \varphi}-{\bf p}_{\varepsilon+n \omega \varphi'}$ 
is the momentum transferred in scattering in the presence of ac field, and ${\bf R}_{\omega}$ 
is a complex vector describing coupling of electron system to this field:
\begin{equation}
{\bf R}_{\omega}=\frac{eE_{\omega}}{\sqrt{2} m \omega} (s_+ + s_-  , (s_+ - s_-)/i ).
%13
\end{equation}
The interaction with phonons is considered under approximation of bulk phonon modes. 
The phonons are characterized by the mode index $\lambda$ and three-dimensional phonon 
momentum ${\bf Q}$ with out-of-plane component $q_z$. The squared matrix element of electron-phonon 
interaction potential is represented as $M_{\lambda {\bf Q}}=C_{\lambda {\bf Q}} I_{q_z}$. 
The squared overlap integral $I_{q_z}=|\left< 0|e^{iq_z z} |0 \right>|^2$ is determined by the 
confinement potential which defines the ground state of 2D electrons, $|0 \rangle$. The 
function $C_{\lambda {\bf Q}}$ characterizes electron-phonon scattering in the bulk.  
The in-plane momenta transferred in electron-phonon collisions, ${\bf q}^{\pm}_{\varepsilon n}$, 
are found from the equation ${\bf q}^{\pm}_{\varepsilon n}={\bf p}_{\varepsilon \varphi}
-{\bf p}_{\varepsilon \pm \omega_{\lambda {\bf Q}^{\pm}} +n \omega \varphi'}$, 
where ${\bf Q}^{\pm}=({\bf q}^{\pm}_{\varepsilon n},q_z)$ and $\omega_{\lambda {\bf Q}}$ 
is the phonon frequency. The effect of the static electric field on the collision integrals 
is given by the energies $\gamma_n={\bf V}_D \cdot {\bf q}_{\varepsilon n}$ and 
$\gamma^{\pm}_n={\bf V}_D \cdot {\bf q}^{\pm}_{\varepsilon n}$, where ${\bf V}_D = 
c [{\bf E} \times {\bf B}]/B^2=(c/B) \hat{\epsilon} {\bf E}$ is the drift velocity 
in the crossed electric and magnetic fields.

In the case of electrons interacting with long-wavelength acoustic phonons in 
cubic lattice, the expressions for $C_{\lambda {\bf Q}}$ and dynamical 
equations needed for determination of $\omega_{\lambda {\bf Q}}$ are the 
following:
\begin{eqnarray}
C_{\lambda {\bf Q}}=\frac{1}{2 \rho_{\scriptscriptstyle M} \omega_{\lambda {\bf Q}}} 
\biggl[ {\cal D}^2 \sum_{ij} {\rm e}_{\lambda {\bf Q} i} {\rm e}_{\lambda {\bf Q} j} q_i q_j  \nonumber \\
+\frac{(eh_{14})^2}{Q^4} \! \! \sum_{ijk,i'j'k'}\kappa_{ijk} \kappa_{i'j'k'} {\rm e}_{\lambda {\bf Q} k} 
{\rm e}_{\lambda {\bf Q} k'} q_i q_j q_{i'} q_{j'} \biggr],
%14
\end{eqnarray}
\begin{equation}
\sum_{j} \left[K_{ij}({\bf Q})  - \delta_{ij} \rho_{\scriptscriptstyle M} \omega^2_{\lambda {\bf Q}} \right] 
{\rm e}_{\lambda {\bf Q} j}=0,
%15
\end{equation}
\begin{eqnarray}
K_{ij}({\bf Q}) =[(c_{11}-c_{44}) q^2_i + c_{44} Q^2] \delta_{ij} \nonumber \\
+ (c_{12}+c_{44}) q_i q_j(1-\delta_{ij}). 
%16
\end{eqnarray}
Here ${\cal D}$ is the deformation potential constant, $h_{14}$ is the piezoelectric 
coupling constant, and $\rho_{\scriptscriptstyle M}$ is the material density. The sums 
are taken over Cartesian coordinate indices. The coefficient $\kappa_{ijk}$ 
is equal to unity if all the indices $i,j,k$ are different and equal to zero 
otherwise. Next, ${\rm e}_{\lambda {\bf Q} i}$ are the components of the unit vector 
of the mode polarization, and $K_{ij}({\bf Q})$ is the dynamical matrix expressed 
through the elastic constants $c_{11}$, $c_{12}$ and $c_{44}$.

Finally, $N_{\lambda {\bf Q}}$ in Eq. (12) is the distribution function of phonons. In the presence 
of thermal gradients this function depends not only on the frequency $\omega_{\lambda {\bf Q}}$, 
but also on the direction of ${\bf Q}$. In the general case, $N_{\lambda {\bf Q}}$ can be 
represented as a sum of symmetric ($s$) and antisymmetric ($a$) parts satisfying the 
relations $N^{s}_{\lambda -{\bf Q}}=N^{s}_{\lambda {\bf Q}}$ and $N^{a}_{\lambda -{\bf Q}}=
-N^{a}_{\lambda {\bf Q}}$, respectively. The drag of electrons by phonons is caused by the 
antisymmetric part. In the linear regime, $N^{a}$ is proportional to $\nabla T$ 
while $N^{s}$ is reduced to the equilibrium distribution function. In particular, one often 
uses the following form:$^{34}$ 
\begin{equation}
N_{\lambda {\bf Q}}= N_{\omega_{\lambda {\bf Q}}} + \frac{\partial N_{\omega_{\lambda {\bf Q}}}}{\partial \omega_{\lambda {\bf Q}}} \frac{\omega_{\lambda {\bf Q}}}{T} \tau_{\lambda} {\bf u}_{\lambda {\bf Q}} 
\cdot \nabla T ,
%17
\end{equation}
obtained from a linearized kinetic equation for phonons in the relaxation time approximation. 
Here $N_{\omega_{\lambda {\bf Q}}}=[\exp(\omega_{\lambda {\bf Q}}/T) -1]^{-1}$ is the equilibrium 
(Planck) distribution function, $\tau_{\lambda}$ is the relaxation time of phonons, and 
${\bf u}_{\lambda {\bf Q}}= \partial \omega_{\lambda {\bf Q}}/\partial {\bf Q}$ is the phonon 
group velocity. Notice that a simple expression ${\bf u}_{\lambda {\bf Q}} = s_{\lambda} {\bf Q}/Q$ 
relating the group velocity to the sound velocity $s_{\lambda}$ is valid only in the isotropic 
approximation. For elastic waves in real cubic crystals the direction 
of ${\bf u}_{\lambda {\bf Q}}$ is not generally coincide with the direction of ${\bf Q}$, 
though the symmetry relation ${\bf u}_{\lambda -{\bf Q}}=-{\bf u}_{\lambda {\bf Q}}$ is
always valid. Substituting Eq. (17) into the expression for the collision integral 
$J^{ph}_{\varepsilon \varphi}$, it is convenient to write the latter as a sum of two parts, 
\begin{equation}
J^{ph}_{\varepsilon \varphi}=J^{ph(0)}_{\varepsilon \varphi} + J^{ph(1)}_{\varepsilon \varphi}, 
%18
\end{equation}
where $J^{ph(0)}_{\varepsilon \varphi}$ contains the equlibrium phonon distribution 
$N_{\omega_{\lambda {\bf Q}}}$ only, while $J^{ph(1)}_{\varepsilon \varphi}$ is determined by 
the anisotropic non-equilibrium correction to phonon distribution [second term in Eq. (17)] 
and is proportional to $\nabla T$. The second term in Eq. (18) is responsible for the phonon drag 
contribution to electric current. 

By using Eqs. (11) and (12), one can directly check the identity $\int d \varepsilon 
D_{\varepsilon} \int d \varphi J_{\varepsilon \varphi}=0$ expressing electron conservation 
requirement. It is worth to emphasize that the collision integrals Eqs. (11) and (12) are written 
in the general form valid for arbitrary relation between radiation frequency $\omega$, phonon 
frequency $\omega_{\lambda {\bf Q}}$ and electron energy $\varepsilon$. In Ref. 23 the 
collision integrals are written in a simpler form valid under the assumptions $\omega \ll 
\varepsilon$ and $\omega_{\lambda {\bf Q}} \ll \varepsilon$. For degenerate electron gas, 
the electrons contributing to electric current have energies $\varepsilon$ close to the Fermi 
energy, and these assumptions usually work very good for microwave frequencies and acoustic 
phonon scattering. However, in the problem of diffusive thermocurrent an extra accuracy 
is required, so the terms of the first order in $\omega/\varepsilon$ are to be retained 
at least in the isotropic part of electron-impurity collision integral (see Eq. (25) below).
 
\section{Solution of kinetic equation}

When searching for the response to temperature gradients only, the effect of the dc 
field in the collision integrals is omitted, $\gamma_n=\gamma^{\pm}_n=0$. It is also 
assumed that the main cause of momentum relaxation of electrons comes from 
electron-impurity scattering rather than from electron-phonon scattering. In GaAs 
quantum wells with electron mobility of about 10$^6$ cm$^2$/V s this approximation 
holds al low temperatures $T < 10$ K (for GaAs quantum well of typical width 20 nm 
the phonon-limited mobility is estimated as $1.3 \times 10^7$ cm$^2$/V s at $T=4.2$ K 
and $3.8 \times 10^6$ cm$^2$/V s at $T=10$ K). Thus, one may neglect the contribution 
$J^{ph(0)}_{\varepsilon \varphi}$ in comparison to $J^{im}_{\varepsilon \varphi}$, 
but the contribution $J^{ph(1)}_{\varepsilon \varphi}$ leading to phonon drag must 
be retained. It is convenient to expand the distribution function in the angular 
harmonics, $f_{\varepsilon \varphi}= \sum_k f_{\varepsilon k} e^{i k \varphi}$. 
The electric current density given by Eq. (6) is determined by the components 
with $k=\pm 1$. Only the effects linear in MW power are considered below. 
The distribution function is represented as a sum of two terms, $f^{(0)}_{\varepsilon k} 
+ f^{(MW)}_{\varepsilon k}$, where $f^{(MW)}_{\varepsilon k}$ is proportional 
to MW power. For $k \neq 0$ the first term is given by the expression 
comprising the diffusive and phonon-drag parts:
\begin{eqnarray} 
f^{(0)}_{\varepsilon k} = -\frac{1}{ik\omega_c+\nu^{(k)} D_{\varepsilon}} \left\{ 
\frac{p_{\varepsilon}}{2m} \left( \frac{\partial f_{\varepsilon k+1} }{\partial T} \nabla_{+} T 
\right.\ \right. \nonumber \\
\left. + \frac{\partial f_{\varepsilon k-1} }{\partial T} \nabla_{-} T \right) 
+\int_0^{2 \pi} \frac{d \varphi}{2 \pi} \int_0^{2 \pi} \frac{d \varphi'}{2 \pi} \sum_{k'} 
e^{i(k'-k)\varphi} \nonumber \\
\left. \times \hat{\cal M} \left\{ \sum_{l=\pm 1} l D_{\varepsilon-l\omega_{\lambda {\bf Q}}}
(f_{\varepsilon-l\omega_{\lambda {\bf Q}} k'} e^{-ik' \theta} - f_{\varepsilon k'} )  
\right\} \right\}, 
%19
\end{eqnarray} 
where $\nabla_{\pm}=\nabla_{x} \pm i \nabla_{y}$, $\nu^{(k)}=\overline{\nu_{\theta}[1-\cos(k\theta)]}$ 
(the line over the expression denotes angular averaging), and $\nu_{\theta} = 
\nu[2p_{\varepsilon}\sin(\theta/2)]$. The integral operator $\hat{\cal M}$ is proportional 
to temperature gradient and defined as
\begin{eqnarray} 
\hat{\cal M} \left\{ A \right\}= 
\sum_{\lambda} \int_{-\infty}^{\infty} \frac{d q_z}{2 \pi} m M_{\lambda {\bf Q}} \tau_{\lambda}
F\left(\frac{\omega_{\lambda {\bf Q}}}{2T} \right) \nonumber \\
\times \left[\frac{1}{Q^2} {\bf q} \cdot \nabla T 
+ \frac{1}{q^2 \omega_{\lambda {\bf Q}} } \frac{\partial \omega_{\lambda {\bf Q}}}{\partial \varphi_q } 
{\bf q} \cdot \hat{\epsilon} \nabla T \right] A, 
%20
\end{eqnarray}
with $F(x)=[x/\sinh(x)]^2$.
It is taken into account that $\omega_{\lambda {\bf Q}} \ll \varepsilon$, which allows one to 
use the quasielastic approximation, when the transferred 2D momentum ${\bf q}^{\pm}_{\varepsilon n}$ 
is replaced by ${\bf q}$, with absolute value $q=2p_{\varepsilon}\sin(\theta/2)$ depending on 
electron energy and scattering angle $\theta=\varphi-\varphi'$. The angle of the vector ${\bf q}$ 
is $\varphi_q=\pi/2+\phi$, where $\phi=(\varphi+\varphi')/2$. The phonon frequency can be written 
as $\omega_{\lambda {\bf Q}}=s_{\lambda {\bf Q}}Q$, where $Q=\sqrt{q^2 + q_z^2}$ and $s_{\lambda 
{\bf Q}}$ is the sound velocity that depends on the mode index and direction of vector ${\bf Q}$. 
If the quantum well is grown in the [001] crystallographic direction, as assumed in the 
following, both $\omega_{\lambda {\bf Q}}$ and $M_{\lambda {\bf Q}}$ are periodic in 
$\varphi_q$ with the period $\pi/2$.    

To find $f^{(MW)}_{\varepsilon k}$ with the accuracy up to the linear terms in MW power, 
only the contributions with low-order, $|n| \leq 1$, Bessel functions $J_n$ are to be 
taken in Eqs. (11) and (12). Physically, this corresponds to a neglect of multi-photon 
absorption processes. If $k \neq 0$, then
\begin{eqnarray} 
f^{(MW)}_{\varepsilon k} \! \! = \frac{P_{\omega}(\varepsilon)/4}{ik\omega_c+\nu^{(k)} D_{\varepsilon}} 
\int_0^{2 \pi} \! \! \frac{d \varphi}{2 \pi}  \int_0^{2 \pi} \! \!
\frac{d \varphi'}{2 \pi} (1-\cos \theta) \nonumber \\
\times [1-b e^{2 i \phi}-b^* e^{-2 i \phi}] \sum_{k'} e^{i(k'-k)\varphi} 
\left. \sum_{n=\pm 1} \right\{ \nu_{\theta} \nonumber \\
\times [ D_{\varepsilon+n\omega} (f_{\varepsilon +n \omega k'} e^{-ik' \theta} \! \! - f_{\varepsilon k'} ) 
- D_{\varepsilon} f_{\varepsilon k'}(e^{-ik' \theta} \! \! -1) ] \nonumber  \\
-\hat{\cal M} \left\{ \sum_{l=\pm 1} l \left[ D_{\varepsilon-l\omega_{\lambda {\bf Q}}+n\omega}
(f_{\varepsilon-l\omega_{\lambda {\bf Q}}+n \omega k'} e^{-ik' \theta} \! \! - f_{\varepsilon k'} ) 
\right. \right. \nonumber \\ 
\left. \left. \left.
-D_{\varepsilon-l\omega_{\lambda {\bf Q}}}(f_{\varepsilon-l\omega_{\lambda {\bf Q}} k'} 
e^{-ik' \theta} - f_{\varepsilon k'} ) \right] \right\} \right\},~~
%21
\end{eqnarray} 
where
\begin{equation}
P_{\omega}(\varepsilon)= \frac{2 e^2 E^2_{\omega} \varepsilon}{m \omega^2} (|s_+|^2+|s_-|^2) 
%22
\end{equation}
is the dimensionless function proportional to MW power (see Eq. (4) for definition of $s_{\pm}$) and 
\begin{equation}
b=\frac{s_- s^*_+}{(|s_+|^2+|s_-|^2)}
%23
\end{equation}
is a complex dimensionless coefficient which depends on the direction of MW polarization and 
determines the sensitivity of transport properties of electrons to this direction. 
The neglect of multi-photon processes implies $P_{\omega}(\varepsilon) \ll 1$. The expression (21) 
comprises both electron-impurity and electron-phonon parts, though only the electron-phonon 
part is essential below.

To find the isotropic ($k = 0$) part of the distribution function, it is necessary to include 
electron-electron scattering into consideration. Though the corresponding collision integral 
$J^{ee}_{\varepsilon}$ is not written in Eq. (5) explicitly, it can be found in Refs. 9 and 23.
The kinetic equation is written as 
\begin{equation} 
J^{im}_{\varepsilon}+ J^{ph(0)}_{\varepsilon}+J^{ee}_{\varepsilon}=0,
%24
\end{equation} 
where only the isotropic part of the distribution function is retained under the 
collision integrals. It is essential that $J^{ee}_{\varepsilon}$ is not affected by 
MW irradiation, while $J^{im}_{\varepsilon}$ is non-zero only in the presence 
of MW irradiation. The distribution $f_{\varepsilon}$ is represented as a sum of 
smooth part $f^{(0)}_{\varepsilon}$ and rapidly oscillating part $f^{(MW)}_{\varepsilon}$.
The smooth part is controlled by electron-electron scattering and, therefore, can be 
approximated by a heated Fermi distribution with effective electron temperature $T_e$, 
the latter is to be found from the energy balance equation 
$\int d \varepsilon D_{\varepsilon} \varepsilon [J^{im}_{\varepsilon}+ J^{ph(0)}_{\varepsilon}]=0$.
For oscillating part, one gets the following expression:
\begin{eqnarray} 
f^{(MW)}_{\varepsilon}= \frac{P_{\omega}(\varepsilon)}{4} \tau_{in} \nu_{tr} 
\sum_{n=\pm 1} \left(1+ \frac{n\omega}{2 \varepsilon} (1-{\cal Z}_{tr}) \right) \nonumber \\
\times
\delta D_{\varepsilon+n\omega} (f_{\varepsilon +n \omega}- f_{\varepsilon}),
%25
\end{eqnarray} 
where $\delta D_{\varepsilon}=D_{\varepsilon}-1$, $\nu_{tr}=\tau_{tr}^{-1}=\nu^{(\pm 1)}$ is 
the transport relaxation rate, and 
\begin{equation}
{\cal Z}_{tr}=\frac{\partial \ln \tau_{tr}}{\partial \ln \varepsilon}
%26
\end{equation}
is the logarithmic derivative of the transport time over energy. The inelastic scattering time $\tau_{in}$ 
entering Eq. (25) describes relaxation of the isotropic oscillating part of electron distribution.$^{9}$
This relaxation is caused mostly by electron-electron scattering and scales with temperature 
as $\tau_{in} \propto T_e^{-2}$. 
  
The electric current is given by Eq. (6), where the distribution function, found from Eqs. (19), (21), 
and (25) with the accuracy up to the terms linear in both $\nabla T$ and $P_{\omega}$, is substituted. 
The thermoelectric tensor $\hat{\beta}$ determining the thermocurrent is represented below as a sum 
of four parts:
\begin{equation}
\hat{\beta}=\hat{\beta}^{(0)}_d + \hat{\beta}^{(0)}_p + \hat{\beta}^{(MW)}_d + \hat{\beta}^{(MW)}_p,
%27
\end{equation}
where diffusive ($d$) and phonon drag ($p$) parts are written separately. Two first terms 
correspond to dark thermocurrent, in the absence of MW irradiation, while the next two 
terms are MW-induced corrections. While $\hat{\beta}^{(0)}_d$ and $\hat{\beta}^{(0)}_p$ are 
determined only by $f^{(0)}_{\varepsilon k}$ from Eq. (19), the MW-induced parts are found 
in a more elaborate way, by combining together the results given by Eqs. (19), (21), and 
(25), as described in Subsection B.   

\subsection{Dark thermocurrent}

In the absence of MWs, the linear response to temperature gradient is found from Eq. (19) 
for $k=\pm 1$ with isotropic ($k'=0$) distribution functions substituted in the right-hand 
side. The thermoelectric coefficients are given by the following expressions: 
\begin{equation}
\hat{\beta}^{(0)}_d= \frac{|e|}{\pi} \int d \varepsilon \frac{\partial f^{(0)}_{\varepsilon}}{\partial T_e} 
\frac{\omega_c \Pi_{\varepsilon} \hat{\epsilon} -\nu_{tr} \varepsilon D^2_{\varepsilon} 
\hat{1} }{\omega_c^2+\nu^2_{tr} D^2_{\varepsilon}}
%28
\end{equation}
and
\begin{eqnarray}
\hat{\beta}^{(0)}_p= \frac{|e|}{2 \pi} \int d \varepsilon 
\frac{\omega_c \varepsilon D_{\varepsilon} \hat{\epsilon} - \nu_{tr} \varepsilon D^2_{\varepsilon} 
\hat{1} }{\omega_c^2+\nu^2_{tr} D^2_{\varepsilon}} \nonumber \\
\times \hat{\cal P}_{1} \left\{ \sum_{l=\pm 1} l D_{\varepsilon-l \omega_{\lambda {\bf Q}}}
(f^{(0)}_{\varepsilon-l\omega_{\lambda {\bf Q}} }  - f^{(0)}_{\varepsilon} ) \omega^{-1}_{\lambda {\bf Q}} 
\right\},
%29
\end{eqnarray}
where $\hat{\cal P}_n$ is the integral operator defined as
\begin{eqnarray}
\hat{\cal P}_n \{A \}=\int_0^{2 \pi} \frac{d \theta}{2 \pi} \int_0^{2 \pi} \frac{d \varphi_q}{2 \pi} 
\sum_{\lambda} \int_{0}^{\infty} \frac{d q_z}{\pi} \nonumber \\
\times (1-\cos \theta)^n
m^2 M_{\lambda {\bf Q}} \tau_{\lambda} F\left(\frac{\omega_{\lambda {\bf Q}}}{2T} \right) 
\frac{2 \omega_{\lambda {\bf Q}}} {Q^2} A.
%30
\end{eqnarray}
The matrices given by Eqs. (28) and (29) contain diagonal symmetric ($\propto \hat{1}$) and 
non-diagonal antisymmetric ($\propto \hat{\epsilon}$) parts, so their symmetry is the same 
as the symmetry of the electrical conductivity. 

The expressions (28) and (29) describe the thermoelectric tensor in a wide region of temperatures and 
magnetic fields. Quantum osicllations of $\hat{\beta}^{(0)}_d$ and $\hat{\beta}^{(0)}_p$ occur 
because of the oscillating dependence of the density of states, $D_{\varepsilon}$. In the 
following, the approximation of overlapping Landau levels is used: $D_{\varepsilon}=1-2d 
\cos(2 \pi \varepsilon/\omega_c)$, where $d = \exp(-\pi/|\omega_c| \tau)$ 
is the Dingle factor ($d \ll 1$) and $\tau$ is the quantum lifetime of electrons, 
given at low temperatures by $\tau=1/\overline{\nu_{\theta}}$. Apart from the condition 
$d \ll 1$, the validity of the expression for $D_{\varepsilon}$ implies $\varepsilon \tau \gg 1$. 
Under the same requirements, $\Pi_{\varepsilon} = \varepsilon - (\omega_c/\pi) d 
\sin(2 \pi \varepsilon/\omega_c)$. The integrals over energy in Eqs. (28) and (29) 
are calculated below under the assumption of strongly degenerate electron gas, and 
the quantum effects up to the second order in the Dingle factors are retained. To take 
into account energy dependence of the Dingle factor due to energy dependence of $\tau$,
the logarithmic derivative ${\cal Z}=\partial \ln \tau/\partial \ln \varepsilon$ is 
introduced. The diffusive part is given by the following expression:
\begin{eqnarray}
\hat{\beta}^{(0)}_d= \frac{\pi |e| T_e}{3(\omega_c^2+\nu^2_{tr}) } 
\left\{ \omega_c \left[ 1 + \frac{2{\cal Z}_{tr} \nu^2_{tr} }{\omega_c^2+\nu^2_{tr}} 
\right. \right. \nonumber\\
\left. \left. + 6 d \cos\frac{2 \pi \varepsilon_F}{\omega_c} \frac{{\cal B}}{X} \right]
\hat{\epsilon} -\nu_{tr}  \left[  1 - {\cal Z}_{tr}  \frac{\omega_c^2 - \nu^2_{tr} }{
\omega_c^2+\nu^2_{tr}}  \right. \right. \nonumber \\
\left. \left. -12 d \frac{\varepsilon_F}{\pi T_e} {\cal B} 
\sin \frac{2 \pi \varepsilon_F}{\omega_c} 
+ 2 d^2 \left(1-{\cal Z}_{tr}  + \frac{2 \pi {\cal Z}}{|\omega_c| \tau}  
\right) 
\right]  \hat{1} \right\}
%31
\end{eqnarray}
with $X=2 \pi^2 T_e/\omega_c$ and ${\cal B}=\partial(X/\sinh X)/\partial X = 
(1-X \coth X)/\sinh X$. All energy-dependent quantities, namely $\nu_{tr}$, $\tau$, ${\cal Z}_{tr}$, 
and ${\cal Z}$, in Eq. (31) are taken at $\varepsilon=\varepsilon_F$. The classical terms 
and the quantum term proportional to $d$ in the diagonal part of $\hat{\beta}^{(0)}_d$ 
have been reported previously.$^{22}$ 

Calculating the phonon-drag part from Eq. (29) under the same approximations, one gets 
the result
\begin{eqnarray}
\hat{\beta}^{(0)}_p= \left. \frac{|e|n_s}{m(\omega_c^2+\nu^2_{tr})} 
\right\{ \omega_c  \left[\Gamma_1 + 2 d^2 \Gamma_{c1} \right] \hat{\epsilon}  \nonumber \\
- \nu_{tr} \left[ \Gamma_1(1 + 2 d^2) + 4 d^2 \Gamma_{c1} \right]  \hat{1} 
\nonumber \\ 
\left. - 4 d \Gamma_{s1}\frac{X}{\sinh X} \cos \frac{2 \pi \varepsilon_F}{\omega_c} 
\left[ \omega_c \hat{\epsilon} - (3/2) \nu_{tr} \hat{1} \right] \right\}.
%32
\end{eqnarray}
Similar to Eq. (31), this expression contains both classical terms and quantum terms 
proportional to $d$ and $d^2$. The dimensionless functions $\Gamma_i$ used here and 
below are defined as 
\begin{eqnarray}
\left( \begin{array}{c} \Gamma_{n} \\ \Gamma_{cn} \\ \Gamma_{sn}  \end{array} \right)= 
\hat{\cal P}^F_n \left\{ \begin{array}{c} 1 \\ 
\cos \frac{2 \pi \omega_{\lambda {\bf Q}}}{\omega_c} \\ 
\frac{\omega_{c}}{2 \pi \omega_{\lambda {\bf Q}}} 
\sin \frac{2 \pi \omega_{\lambda {\bf Q}}}{\omega_c} \end{array} \right\},
%33
\end{eqnarray}
where $\hat{\cal P}^F_n$ denotes $\hat{\cal P}_n$ at $\varepsilon=\varepsilon_F$. 
The function $\Gamma_{1}$ determines the {\em classical} contribution$^{34}$ to 
phonon-drag thermoelectric response and does not depend on the magnetic field. This contribution has 
been considered previously in the isotropic approximation for phonon spectrum, when there 
are one longitudinal phonon branch with velocity $s_l$ and two transverse 
branches with velocity $s_t$. For high temperatures, when $T$ exceeds both 
$s_{\lambda}p_F$ and $\pi s_{\lambda}/a$ ($p_F$ is the Fermi momentum and $a$ 
is the quantum well width), $\Gamma_{1}$ is temperature-independent. At low 
temperatures, $T \ll s_{\lambda}p_F$, the Bloch-Gruneisen transport regime 
is realized, when electron scattering by phonons occurs at small angles, 
$\theta \ll 1$. In this regime,$^{35}$ $\Gamma_{1}$ scales with temperature as 
$T^2$ (or as $T^4$ if only the deformation-potential mechanism of electron-phonon 
interaction is present). The functions $\Gamma_{cn}$ and $\Gamma_{sn}$ standing in 
the quantum contributions depend on the magnetic field and can be analytically 
calculated only in certain limits (see Appendix).

The terms $\propto d$ in Eqs. (31) and (32) describe the Shubnikov-de Haas oscillations 
of the thermocurrent. The forthcoming consideration, however, is focused at the case of 
$|\omega_c| \ll 2 \pi^2 T_e$, which means that $X/\sinh X$ is exponentially small so the 
Shubnikov-de Haas oscillations are suppressed and the quantum corrections are given by 
the terms $\propto d^2$ only. Using $n_s=m\varepsilon_F/\pi$, one may check that the 
tensor (31) under these conditions satisfies the Mott relation $\hat{\beta}^{(0)}_d=
-(\pi^2 T_e/3|e|) (\partial \hat{\sigma}/\partial \varepsilon_F)$, where $\hat{\sigma}= 
\hat{1} \sigma_{d} - \hat{\epsilon} \sigma_{\bot}$ is the conductivity tensor whose components 
are $\sigma_{d}=e^2 n_s \nu_{tr}(1 + 2 d^2)/[m(\omega_c^2+\nu_{tr}^2)]$ and $\sigma_{\bot}=e^2 n_s 
\omega_c/[m( \omega_c^2+\nu_{tr}^2)]$. The quantum corrections both in these 
expressions and in Eq. (31) are essential only in the classically strong magnetic fields, 
so the terms $\propto (\nu_{tr}/\omega_c)d^2$ are neglected in comparison to the terms 
$\propto d^2$. 

The diffusive thermoelectric coefficients do not oscillate before the onset of Shubnikov-de Haas 
oscillations. In contrast, quantum magnetooscillations of phonon-drag thermoelectric coefficients 
persist under the assumed condition $|\omega_c| \ll 2 \pi^2 T_e$, because of the 
presence of $\Gamma_{c1}$ in $\hat{\beta}^{(0)}_p$. Indeed, the oscillating nature 
of the function $\cos (2 \pi \omega_{\lambda {\bf Q}}/\omega_c)$ is not completely 
washed out after the integration under $\hat{\cal P}$. The major contribution to such 
integrals comes from the region of variables around $q_z = 0$ and $\theta = \pi$, 
which physically corresponds to backscattering of electrons as a result of emission or 
absorption of phonons moving in the quantum well plane, the wavenumber of these
phonons is close to $2p_F$. Thus, there exist resonant phonon frequencies, roughly 
estimated as $2 p_F s_{\lambda}$, which lead to magnetooscillations of phonon-drag 
thermopower observed$^{20}$ in high-mobility samples. With decreasing temperature, 
the oscillations are exponentially suppressed in the Bloch-Gruneisen regime 
(see Appendix). The same kind of oscillations is observed in electrical resistivity, 
they are known as acoustic magnetophonon oscillations or phonon-induced 
resistance oscillations.$^{24-29}$  

\subsection{Microwave-induced thermocurrent}

The distribution functions $f_{\varepsilon 1}$ and $f_{\varepsilon -1}$ determining 
the electric current under MW irradiation are to be found up to the terms linear in 
$P_{\omega}(\varepsilon)$. There are two types of such MW-induced contributions. The direct 
ones are obtained in two ways: i) by calculating $f^{(MW)}_{\varepsilon \pm 1}$ from Eq. (21), 
where the isotropic distribution function $f^{(0)}_{\varepsilon}$ is retained under the integral 
(only the phonon part is essential), and ii) by calculating $f^{(0)}_{\varepsilon \pm 1}$ 
from Eq. (19), where the isotropic MW-induced distribution function $f^{(MW)}_{\varepsilon}$ 
is placed in the right-hand side. The indirect contributions assume calculation of 
$f^{(0)}_{\varepsilon \pm 1}$ and $f^{(MW)}_{\varepsilon \pm 1}$ by substituting 
anisotropic parts of $f^{(MW)}_{\varepsilon k}$ and $f^{(0)}_{\varepsilon k}$, 
respectively, in the right-hand sides of Eq. (19) and Eq. (21). Similar technique 
has been used for calculation of the MW-induced conductivity. Following the 
notations of Ref. 10, one may denote the direct contributions (i) and (ii) as the 
"displacement" and "inelastic" ones, respectively, and the indirect contributions 
as the "quadrupole" ones. Strictly speaking, there exists one more indirect 
contribution called the "photovoltaic" one,$^{10}$ which is determined by the 
MW-generated time-dependent part of the distribution function and cannot be obtained 
from the kinetic equation Eq. (5) because the latter is written for time-independent 
$f_{\varepsilon \varphi}$. The indirect contributions to $\hat{\beta}$ begin with 
the terms of the order $\nu_{tr}/\omega_c$ compared to direct contributions. Since 
all the MW-induced contributions are of quantum nature and important only in the 
region of classically strong magnetic field, $\omega_c \gg \nu_{tr}$, the indirect 
contributions are less significant than the direct ones and can be safely 
neglected in the thermopower coefficients presented in the next section. 
Therefore, the attention below is focused at the direct contributions only. 

The current is calculated in the regime when electron gas is degenerate. 
Within the required accuracy, the solution of Eq. (25) is given by the 
following expression:
\begin{eqnarray}
f^{(MW)}_{\varepsilon}= \frac{d}{2} P_{\omega}(\varepsilon) \tau_{in} \nu_{tr} \left\{
\sin \frac{2 \pi \varepsilon}{\omega_c} \sin \frac{2 \pi \omega}{\omega_c} \right. \nonumber \\ 
\left. \times (f^{(0)}_{\varepsilon+\omega}-f^{(0)}_{\varepsilon-\omega} ) - 
\cos \frac{2 \pi \varepsilon}{\omega_c} \cos \frac{2 \pi \omega}{\omega_c} \right. \nonumber \\
\left. 
\times \left[f^{(0)}_{\varepsilon+\omega}+f^{(0)}_{\varepsilon-\omega} -2 f^{(0)}_{\varepsilon} 
\right. \right. \nonumber \\
\left. \left. +\frac{\omega}{2 \varepsilon} \left(1-{\cal Z}_{tr} + \frac{2 \pi {\cal Z}}{
|\omega_c|\tau} \right) (f^{(0)}_{\varepsilon+\omega}-f^{(0)}_{\varepsilon-\omega} ) 
\right] \right\}.
%34
\end{eqnarray}
The first term of this expression gives the main contribution sufficient for calculation 
of the MW-induced resistance.$^9$ The second term represents a correction of the order 
$\omega/\varepsilon$, which is necessary for calculation of the diffusive thermopower. 
The term proportional to the factor $\sin(2 \pi \omega/\omega_c)$ in Eq. (34) 
also enters Eqs. (35), (37), (43), (45), and (48) below, where the contribution of 
inelastic mechanism is present. This factor reflects the property$^{9}$ that the strongest 
modification of the electron distribution function under MW irradiation in the presence of 
weak Landau quantization occurs when $\omega/\omega_c=n \pm 1/4$ ($n$ is an integer). 
The correction proportional to the factor $\cos(2 \pi \omega/\omega_c)$ appears because 
the resonance absorption of MW radiation at $\omega/\omega_c=n$ also has an effect on the 
distribution function. 

The consideration below assumes the approximation $|\omega_c| \ll 2 \pi^2 T_e$, when
Shubnikov-de Haas oscillations are thermally averaged out. This dramatically simplifies 
calculation of the integrals over energy because one can average the products 
of rapidly oscillating functions such as $D_{\varepsilon}$ and $f^{(MW)}_{\varepsilon}$ 
over the period $\omega_c$ before integration over the energy. After substituting Eq. (34) 
into the first part of the right-hand side of Eq. (19) and calculating the current 
according to Eq. (6) (one may equally use Eq. (28) with $f^{(0)}_{\varepsilon}$ replaced 
by $f^{(MW)}_{\varepsilon}$), the diffusive part of $\hat{\beta}$ takes the form
\begin{eqnarray}
\hat{\beta}^{(MW)}_d= \frac{|e| d^2 \tau_{in} \nu_{tr} \omega^2 P_{\omega} {\cal T}_{in}}{\pi T_e 
(\omega_c^2+\nu^2_{tr}) } \nonumber \\ 
\times \left[ \frac{\omega_c^2}{2 \pi \omega} \sin \frac{2 \pi \omega }{\omega_c} 
\hat{\epsilon}  -\nu_{tr} (1-{\cal Z}_{tr})\cos \frac{2 \pi \omega }{\omega_c}  
\hat{1} \right],
%35
\end{eqnarray}
where all energy-dependent quantities are taken at $\varepsilon= \varepsilon_F$, in 
particular, $P_{\omega} \equiv P_{\omega}(\varepsilon_F)$. The main contribution to 
the derivative over temperature in Eq. (19) comes from temperature dependence of 
the inelastic relaxation time, expressed through the logarithmic derivative 
\begin{equation}
{\cal T}_{in}= \frac{\partial \ln \tau_{in}}{\partial \ln T_e} \simeq -2. 
%36
\end{equation}
The factor $\sin(2 \pi \omega/\omega_c)$ typical for MW-induced conductivity$^9$ does 
not appear in the diagonal part of thermoelectric tensor Eq. (35), because of different 
dependence of the diffusive thermoelectric current on electron energy distribution as 
compared to the drift current. The first term of $f^{(MW)}_{\varepsilon}$ is averaged 
out in the diagonal components of $\hat{\beta}^{(MW)}_d$, while the second term of 
$f^{(MW)}_{\varepsilon}$, proportional to $\cos(2 \pi \omega/\omega_c)$, survives 
this averaging.  

For phonon-drag part of $\hat{\beta}$ the result is the following: 
\begin{eqnarray}
\hat{\beta}^{(MW)}_p=\frac{2|e|n_s d^2 P_{\omega} }{m (\omega^2_c+\nu_{tr}^2)} \left\{ \nu_{tr} 
\tau_{in} \Gamma_{s1} \frac{2 \pi \omega}{\omega_c} \sin \frac{2 \pi \omega}{\omega_c} \right. \nonumber \\
\times [-\omega_c \hat{\epsilon}+(3/2)\nu_{tr} \hat{1}]   \nonumber \\
+ \left( \Gamma_{c2} \sin^2 \frac{\pi \omega }{\omega_c} + \Gamma_{s2} 
\frac{\pi \omega}{\omega_c} \sin \frac{2 \pi \omega }{\omega_c}  \right) \nonumber \\ 
\times [\omega_c (-\hat{\epsilon} + \hat{g}_0) + 2 \nu_{tr} (\hat{1}+ \hat{h}_0 )] \nonumber \\
+ \left( \tilde{\Gamma}_{c2} \sin^2 \frac{\pi \omega }{\omega_c} + \tilde{\Gamma}_{s2} 
\frac{\pi \omega}{\omega_c} \sin \frac{2 \pi \omega }{\omega_c}  \right) \nonumber \\ 
\times \left. [\omega_c \hat{g}_1 + 2 \nu_{tr} \hat{h}_1]
\right\},
%37
\end{eqnarray}
where 
\begin{eqnarray}
\hat{g}_0= b' \hat{\sigma}_x + b'' \hat{\sigma}_z,~~
\hat{g}_1= b' \hat{\sigma}_x - b'' \hat{\sigma}_z, \nonumber \\
\hat{h}_0= b' \hat{\sigma}_z - b'' \hat{\sigma}_x,~~
\hat{h}_1= b' \hat{\sigma}_z + b'' \hat{\sigma}_x, 
%38
\end{eqnarray}
$\hat{\sigma}_z=\left( \begin{array}{cc} 1 & 0 \\ 0 & -1 \end{array} \right)$ and 
$\hat{\sigma}_x=\left( \begin{array}{cc} 0 & 1 \\ 1 & 0 \end{array} \right)$ are the 
Pauli matrices, while $b'$ and $b''$ denote real and imaginary parts of $b$ [see Eq. (23)], 
respectively. The quantities $\tilde{\Gamma}_{i}$ differ from ${\Gamma}_{i}$ by placing the factor
\begin{eqnarray}
\cos(4 \varphi_q)- \sin(4 \varphi_q)\left(1+ \frac{q^2_z}{4 p^2_{\varepsilon} \sin^2 \theta/2} \right) \frac{1}{\omega_{\lambda {\bf Q}} } 
\frac{\partial \omega_{\lambda {\bf Q}}}{\partial \varphi_q} 
%39
\end{eqnarray}
under the integral operator $\hat{\cal P}$ in Eq. (33). In the general case, both terms 
in Eq. (39) are essential for calculation of $\tilde{\Gamma}_{i}$. In the isotropic 
approximation for phonon spectrum the second term in Eq. (39) vanishes, but $\tilde{\Gamma}_{i}$ 
is still nonzero, because the piezoelectric-potential part of $M_{\lambda {\bf Q}}$ 
remains angular-dependent. If the anisotropy of phonon spectrum is weak, the second 
term in Eq. (39) can be neglected in calculation of piezoelectric potential contribution.

The expression (37) includes contributions from both inelastic (first term) and displacement 
(second and third terms) mechanisms. The inelastic-mechanism contribution can be obtained from 
Eq. (29) after replacing $f^{(0)}_{\varepsilon}$ by $f^{(MW)}_{\varepsilon}$.
The displacement-mechanism contribution has the form similar to that of MW-induced contribution to conductivity,$^{8}$ as it contains the factors $\sin(2 \pi \omega/\omega_c)$ and $\sin^2(\pi 
\omega/\omega_c)$. The first of these factors has extrema at $\omega/\omega_c=n \pm 1/4$, corresponding 
to the conditions of maximal displacement of electrons along the effective drag force or against 
this force under photon-assisted scattering, similar as in the case of a response to 
dc field.$^7$ The second factor describes the enhancement of photon-assisted scattering 
probabilities in the resonance, $\omega/\omega_c=n$, and their suppression in the 
anti-resonance, $\omega/\omega_c=n+1/2$. The displacement-mechanism contribution 
depend on MW polarization direction through the terms with the matrices of Eq. (38). 

The fundamental difference between $\hat{\beta}^{(MW)}_p$ and the MW-induced contribution to 
conductivity is given by the factors $\Gamma_{i}$ and $\tilde{\Gamma}_{i}$, which are not merely 
constants but functions of the magnetic field describing the magnetophonon oscillations. The products 
of these magnetophonon oscillating factors by the MW-induced oscillating factors $\sin(2 \pi 
\omega/\omega_c)$ and $\sin^2(\pi \omega/\omega_c)$ physically correspond to the interference 
of these two kinds of oscillations and can be viewed as a result of photon and phonon frequency 
mixing in the scattering probabilities.

The phonon-drag part of $\hat{\beta}$ depends on electron temperature $T_e$ through 
the inelastic scattering time $\tau_{in}$. The quantities ${\Gamma}_{i}$ and 
$\tilde{\Gamma}_{i}$ are determined by the lattice temperature $T$.

There is an important question of whether the thermoelectric tensor $\hat{\beta}$  
satisfies the symmetry with respect to time inversion (Onsager symmetry). In the absence 
of microwaves, this symmetry, of course, is satisfied. Under MW irradiation, when electrons 
are out of equilibrium, the Onsager symmetry can be broken.$^{10}$ In application to 
the problem of electrons in the presence of electromagnetic waves, the time inversion 
implies, apart from the magnetic field reversal $\omega_c \rightarrow -\omega_c$, the 
transformations ${\bf e} \rightarrow {\bf e}^*$ and ${\bf k} \rightarrow -{\bf k}$, 
where ${\bf k}$ is the wave vector of the electromagnetic wave. ${\bf e} \rightarrow 
{\bf e}^*$ means that ${\rm e}_{\pm} \rightarrow {\rm e}_{\mp}^*$, which is equivalent 
to $\kappa_{\pm} \rightarrow \kappa_{\mp}$ (see the beginning of Sec. II), while 
${\bf k} \rightarrow -{\bf k}$ means that the sign at $\omega_p$ in Eq. (4) for 
$s_{\pm}$ is inverted, as follows from reversibility of the wave transmission problem$^{30}$ 
employed for derivation of Eq. (3). Therefore, the denominator in Eq. (4) 
transforms as $\omega \pm \omega_c + i \omega_p 
\rightarrow \omega \mp \omega_c - i \omega_p$, which results in $s_{\pm} 
\rightarrow s^*_{\mp}$ under the time inversion. The Onsager symmetry relation 
takes the form
\begin{equation}
\beta_{ij}(\omega_c,s_-,s_+)=\beta_{ji}(-\omega_c,s_+^*,s_-^*),
%40
\end{equation}
and similar relations can be written for the other transport coefficients including the 
conductivity. From $s_{\pm} \rightarrow s^*_{\mp}$ one can 
see that both $|s_+|^2+|s_-|^2$ and $s_-s_+^*$ are invariants with respect to time 
inversion, thus the function $b$ defined by Eq. (23) is also an invariant. The MW-induced 
part of $\hat{\beta}$ given by Eq. (37) does contain the terms violating the Onsager
symmetry Eq. (40), these are the terms at the matrices $\hat{g}_0$ and $\hat{g}_1$. These 
terms change their signs under magnetic field reversal but are invariant under permutation 
of Cartesian indices. 

\section{Thermopower coefficients}

Having found $\hat{\beta}$, one may calculate the thermopower tensor $\hat{\alpha}$, 
which is presented below as a sum of dark and MW-induced parts, $\hat{\alpha}=
\hat{\alpha}^{(0)}+ \hat{\alpha}^{(MW)}$. Because of the presence of terms which 
depend on MW polarization, this tensor is a general matrix.
        
In the absence of MW irradiation, $\hat{\alpha}$ has the same symmetries as the 
resistivity tensor, $\alpha_{xx}^{(0)}=\alpha_{yy}^{(0)}$ and $\alpha_{xy}^{(0)}=
-\alpha_{yx}^{(0)}$. By using the expressions $\rho^{(0)}_{xy} =m \omega_c/e^2n_s$ 
and $\rho^{(0)}_{xx}= m \nu_{tr}(1+2d^2)/e^2n_s$ together with Eqs. (31) and (32), 
where the Shubnikov-de Haas terms are neglected, one obtains, within the accuracy 
up to $d^2$, the following results:  
\begin{eqnarray}
\alpha^{(0)}_{xx}
=-\frac{\pi^2}{3 |e|} \frac{T_e}{\varepsilon_F} 
\left(1+ \frac{{\cal Z}_{tr} \nu^2_{tr} }{
\omega_c^2+\nu^2_{tr}} \right) \nonumber \\
- \frac{1}{|e|}(\Gamma_1+2d^2 \Gamma_{c1}),
%41
\end{eqnarray}
\begin{eqnarray}
\alpha^{(0)}_{xy}
=\frac{\omega_c \nu_{tr} }{\omega_c^2+\nu^2_{tr}} \left[ \frac{\pi^2}{3 |e|} \frac{T_e}{\varepsilon_F} \right. \nonumber \\
\left. \times \left({\cal Z}_{tr}+2d^2\left({\cal Z}_{tr}-\frac{2 \pi {\cal Z}}{|\omega_c| \tau} \right) \right)
-\frac{1}{|e|} 2 d^2 \Gamma_{c1} \right].
%42
\end{eqnarray}
In the classical case, the thermopower coefficients have usual forms found in 
literature.$^{22}$ The Landau quantization leads to additional terms proportional 
to $d^2$. In the phonon-drag part of thermopower, these terms are determined by the 
function $\Gamma_{c1}$ oscillating with the magnetic field. Because of these quantum 
corrections, the transverse phonon-drag thermopower is nonzero. In Fig. 1 
the longitudinal and transverse thermopower are plotted as functions of magnetic 
field for a rectangular GaAs quantum well of width 14 nm, with electron 
density $n_s=5 \times 10^{11}$ cm$^{-2}$ and mobility $2 \times 10^6$ cm$^2$/V s. 
The quantum lifetime $\tau=7$ ps is assumed, which corresponds to the 
ratio $\tau_{tr}/\tau \simeq 11$. The 
phonon scattering time $\tau_{\lambda}$ is chosen as 0.2 $\mu$s for each mode, which approximately 
corresponds to 1 mm mean free path for phonons.$^{36}$ The elastic coefficients for 
GaAs in units 10$^{11}$ dyn/cm$^2$ are $c_{11}=12.17$, $c_{12}=5.46$, and $c_{44}=6.16$. 
The deformation potential, piezoelectric coefficient, and density are 
${\cal D}=7.17$ eV, $h_{14}=1.2$ V/nm, and $\rho=5.317$ g/cm$^3$, respectively.
The energy dependence of the transport time and quantum lifetime is assumed to be 
$\propto \varepsilon^{3/2}$ and $\propto \varepsilon^{1/2}$, respectively, which 
corresponds to $\nu(q) \propto \exp(-l_c q)$ under the condition of small-angle scattering, 
when $l_c p_F \gg 1$. The oscillations of the thermopower coefficients are caused by 
magnetophonon resonances. At low temperature, the oscillations are barely visible because 
the system falls into the Bloch-Gruneisen regime, but they are essential at higher temperatures. 
The last peak of $\alpha^{(0)}_{xx}$ is due to the scattering of electrons by high-energy 
(longitudinal) phonons, this peak disappears first with lowering temperature. 
The non-oscillating, proportional to $1/B$, part of $\alpha^{(0)}_{xy}$ is 
determined by the diffusive contribution.

\begin{figure}[ht]
\includegraphics[width=9.2cm]{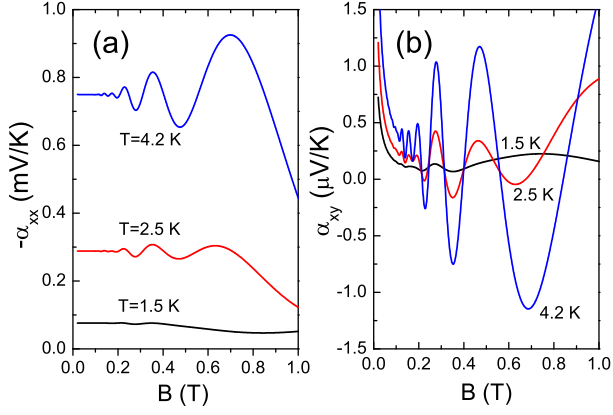}
\caption{(Color online) Longitudinal (a) and transverse (b) thermopower in the absence 
of microwave excitation plotted for three temperatures. The calculations are done for 
a GaAs quantum well of width 14 nm with electron density $n_s=5 \times 10^{11}$ cm$^{-2}$ 
and mobility $2 \times 10^6$ cm$^2$/V s. The quantum lifetime of electrons is $\tau=7$ ps.}  
\end{figure}

Let us consider now the thermopower coefficients in the presence of MW excitation. 
While Eqs. (41) and (42) are valid for both classically strong and classically weak 
magnetic fields, the MW-induced contributions are important only in the limit of 
classically strong magnetic fields. For this reason, only a part of the terms 
presented in Eqs. (35) and (37) are essential for calculation 
of thermopower in this limit. In particular, the longitudinal thermopower in 
classically strong magnetic fields is written simply as $\alpha_{xx}=\rho_{xy} 
\beta_{yx}$. The influence of microwaves on Hall resistivity $\rho_{xy}$ is 
weak$^{12}$, so $\alpha_{xx}$ is directly determined by $\beta_{yx}$. Neglecting 
the contributions of higher order in $\nu_{tr}/\omega_c$ in Eqs. (35) and (37), 
one obtains
\begin{eqnarray}
\alpha^{(MW)}_{xx}= \frac{2 d^2 P_{\omega}}{|e|} 
\left\{ \nu_{tr} \tau_{in} \frac{2 \pi \omega}{\omega_c}  \sin \frac{2 \pi \omega}{\omega_c} 
\right. \nonumber \\
\times \left( \Gamma_{s1} -  \frac{{\cal T}_{in} \omega_c^2}{8 \pi^2 \varepsilon_F T_e} 
\right) \nonumber \\ 
+ [(1+b')\Gamma_{c2}+b'\tilde{\Gamma}_{c2}] \sin^2 \frac{\pi \omega}{\omega_c} 
\nonumber \\
\left.
+[(1+b')\Gamma_{s2}+b'\tilde{\Gamma}_{s2}] \frac{\pi \omega}{\omega_c} \sin
\frac{2 \pi \omega}{\omega_c} \right\},
%43
\end{eqnarray}
while $\alpha^{(MW)}_{yy}$ differs from this expression by changing the sign at $b'$.
The first term in Eq. (43) is caused by modification of the isotropic distribution 
function of electrons by microwaves (inelastic mechanism) and includes both the 
phonon-drag and the diffusive contributions. Since the diffusive term increases 
with decreasing temperature, it may become comparable to the phonon-drag one. However, 
inevitable heating of electron gas by microwaves tends to hinder the contribution of the 
diffusive term. The remaining terms in Eq. (43) describe the phonon-drag thermopower 
caused by the displacement mechanism. They contain contributions 
proportional to $b'$, which change the symmetry of the thermopower coefficients. The 
dependence of these contributions on the polarization angle $\chi$ can be illustrated 
for the case of linear polarization of the incident wave, when $b$ is represented 
in the form
\begin{equation}
b=\frac{1}{2} e^{-2 i \chi}\frac{\omega^2-\omega_c^2+\omega^2_p-2 i \omega_c \omega_p}{
\omega^2+\omega_c^2+\omega^2_p}.
%44
\end{equation}
Since $b' \equiv {\rm Re}(b)$ contains the terms both even and odd in magnetic field, 
$\alpha_{xx}$, in general, is not symmetric in $B$ (the reversal of magnetic 
field means alteration of the sign of $\omega_c$ in all equations). 
The "inelastic" contribution in Eq. (43) should dominate at low enough temperatures, 
when $\nu_{tr} \tau_{in} > 1$. The "displacement" terms become more important with 
increasing temperature. It is worth to emphasize that the oscillations in these terms 
due to the factor $\sin^2(\pi \omega/\omega_c)$ are comparable by amplitude with 
the oscillations due to the factor $\sin(2 \pi \omega/\omega_c)$. This behavior is 
in contrast with that for MW-induced resistance. In the resistance, the contribution
at $\sin(2 \pi \omega/\omega_c)$ dominates because it overcomes the oscillating part 
of $\sin^2(\pi \omega/\omega_c)$ by the factor $2 \pi \omega/\omega_c$ which is 
numerically large in the region $\omega > \omega_c$ where MIRO are observed. As a 
consequence, the MW-induced resistance magnetooscillations due to the displacement 
mechanism are very similar to the magnetooscillations due to the inelastic 
mechanism,$^9$ so these two mechanisms are difficult to separate experimentally. 
In the phonon-drag thermopower, the contributions at $\sin(2 \pi \omega/\omega_c)$ 
and $\sin^2(\pi \omega/\omega_c)$ are proportional to the functions $\Gamma_{s2}$ 
and $\Gamma_{c2}$, respectively, and $\Gamma_{c2}$ is larger than $\Gamma_{s2}$. 
Moreover, $\Gamma_{c2} \gg \Gamma_{s2}$ in the region of low magnetic fields, 
$|\omega_c| \ll 4 \pi p_F s_{\lambda}$, see the Appendix. The ratio of the amplitudes 
of $\sin(2 \pi \omega/\omega_c)$ and $\sin^2(\pi \omega/\omega_c)$ oscillations in 
the "displacement" part of the thermopower is estimated as $\omega/2p_F s_{\lambda}$, 
which is of the order of unity for typical electron densities and MW frequencies. 
The same is true for the "displacement" contribution to transverse thermopower 
described below by Eq. (47).
   
A more careful analysis is required for evaluation of the transverse (Nernst-Ettingshausen) 
thermopower, because the latter is determined by both diagonal and non-diagonal parts 
of $\hat{\beta}$ and is sensitive to MW-induced modifications of the longitudinal 
resistivity. Indeed, $\alpha_{xy}=\rho_{xy}\beta_{yy}+\rho_{xx}\beta_{xy}$. 
The influence of microwaves on $\rho_{xy}$ is weak and not essential for 
determination of $\alpha_{xy}$, while their influence on $\rho_{xx}$ 
is strong. Under the assumed condition that the electron-impurity 
scattering is more important than electron-phonon scattering, the 
longitudinal resistivity correction due to MW irradiation is written as$^9$
\begin{eqnarray}
\rho^{(MW)}_{xx}= - \frac{2d^2 m \nu_{tr}^2 \tau_{in}}{e^2 n_s} 
P_{\omega} \frac{2 \pi \omega}{\omega_c} \sin \frac{2 \pi \omega}{\omega_c},
%45
\end{eqnarray}
and $\rho^{(MW)}_{yy}=\rho^{(MW)}_{xx}$. Equation (45) implies that $\rho^{(MW)}_{xx}$ 
is governed by the inelastic mechanism. The displacement mechanism for electron-impurity 
scattering is less important at low temperatures, especially in the case of small-angle 
scattering processes relevant for high-mobility 2D systems.$^9$ In contrast, for 
electron-phonon scattering determining phonon-drag thermopower, the displacement 
mechanism is significant under the condition $\omega_{\lambda {\bf Q}} \leq 2 T$ 
when the main contribution to oscillating functions $\Gamma_{c2}$ and $\Gamma_{s2}$ 
comes from large-angle scattering processes (backscattering). Among the "displacement" 
terms contributing into the transverse 
thermopower $\alpha^{(MW)}_{xy}$ there is a strong polarization-dependent term coming 
from the diagonal part of the matrices $\hat{g}_0$ and $\hat{g}_1$ in Eq. (37). The other 
contributions to $\alpha^{(MW)}_{xy}$ contain a small factor $\nu_{tr}/\omega_c$.  
Out of them, only the "inelastic" ones can compete with the mentioned 
polarization-dependent contribution. Therefore, with the assumed accuracy up to $d^2$,  
the result is written as a sum of two terms: 
\begin{eqnarray}
\alpha^{(MW)}_{xy} \simeq \Delta \alpha_{xy} \sin(2 \chi+\eta_{{\bf B}}) + \alpha^{in}_{xy}, 
%46
\end{eqnarray}
where
\begin{eqnarray}
\Delta \alpha_{xy} = \frac{2 d^2 P_{\omega}}{|e|} |b| 
\left[ (\Gamma_{c2}- \tilde{\Gamma}_{c2}) \sin^2 \frac{\pi \omega}{\omega_c} \right. \nonumber \\
\left. + (\Gamma_{s2}- \tilde{\Gamma}_{s2}) \frac{\pi \omega}{\omega_c}  
\sin \frac{2 \pi \omega}{\omega_c}  \right],
%47
\end{eqnarray}
and
\begin{eqnarray}
\alpha^{in}_{xy} = - \frac{2 d^2 P_{\omega} \nu_{tr}^2 \tau_{in}}{|e|\omega_c} \nonumber \\
\times \left[ \frac{2 \pi \omega}{\omega_c} 
\sin \frac{2 \pi \omega}{\omega_c} \left(\Gamma_{1} + \frac{\pi^2 T_e}{3 \varepsilon_F} 
-\frac{\Gamma_{s1}}{2} \right) \right. \nonumber \\
\left.  - \frac{\omega^2 {\cal T}_{in} }{2 T_e \varepsilon_F} 
\left( \frac{\omega_c}{2 \pi \omega} \sin \frac{2 \pi \omega}{\omega_c} 
- (1-{\cal Z}_{tr}) \cos \frac{2 \pi \omega}{\omega_c} \right) \right].
%48
\end{eqnarray}
To obtain $\alpha^{(MW)}_{yx}$, one should change the sign at the second 
term in Eq. (46). Since the effects under consideration are linear in 
MW intensity, the polarization-dependent term is a harmonic 
function of the doubled polarization angle; a similar angular dependence 
is expected for electrical resistivity.$^{37}$
This term is characterized by the amplitude $\Delta \alpha_{xy}$ and the 
phase angle $\eta_{{\bf B}}$ which are, respectively, a symmetric and 
an antisymmetric function of the magnetic field. For linear polarization, 
when Eq. (44) is valid, the phase angle is defined as $\tan \eta_{{\bf B}} = 
2 \omega_c \omega_p/(\omega^2-\omega_c^2+\omega_p^2)$. One may introduce 
the effective polarization angle $\chi_{{\bf B}}=\chi+\eta_{{\bf B}}/2$ 
describing the direction of the ac electric field in the 2D plane, which 
is different from the polarization of the incident wave.
The polarization-dependent term, in general, is not antisymmetric under 
reversal of ${\bf B}$, though for special orientation of the incident ac 
field along $x$ or $y$ axes the symmetry property $\alpha^{(MW)}_{xy}({\bf B}) = 
- \alpha^{(MW)}_{xy}(-{\bf B})$ is preserved. If the angle 
$\chi_{{\bf B}}$ is equal to $\pi/2$ or $0$, which means that the 
electric field in the 2D plane is polarized along $y$ or $x$ axes (i.e. along 
or perpendicular to the temperature gradient), the polarization-dependent term 
is equal to zero. The contribution of this term can be experimentally 
distinguished from the other contributions by its dependence on the 
polarization.

The polarization-independent term given by Eq. (48) contains several contributions 
of different origin, though all of them are caused by the inelastic mechanism. The 
first part [the second line of Eq. (48)] comprises three different contributions.
The first one, at $\Gamma_1$, comes from the MW-induced correction to resistance 
if the thermoelectric current is due to the phonon-drag mechanism. 
The second contribution comes from the MW-induced correction to resistance 
if the thermoelectric current is due to the diffusive mechanism. These 
two contributions can be distinguished from each other by their temperature 
dependence. At low temperatures (roughly estimated as $T_e < 0.5$ K), the second 
contribution can exceed the first one, as it decreases with $T_e$ slower 
[see Eq. (A11) for low-temperature behavior of $\Gamma_1$]. However, the MW 
heating of electron gas renders this regime practically unrealizable. The third 
contribution, at $\Gamma_{s1}$, is caused by the MW-induced correction to
the phonon-drag part of thermoelectric tensor. In contrast to the first and 
second contributions, this one contains magnetophonon oscillations. However, 
in the region of fields where these oscillations exist, $|\omega_c| < 2 p_F 
s_{\lambda}$, the term $\Gamma_{s1}/2$ is much smaller than $\Gamma_1$. 
The second part [the last line of Eq. (48)] contains the contributions due to 
MW-induced correction to diffusive part of thermoelectric tensor. This part 
does not exceed the contribution proportional to $\pi^2 T_e/3 \varepsilon_F$ in the 
second line of Eq. (48) under the assumed comdition $|\omega_c| \ll 2 \pi^2 T_e$. 
Therefore, the contribution proportional to $\Gamma_{1}$ dominates over the others 
in Eq. (48) in the relevant region of parameters. This means that magnetooscillations 
of $\alpha^{in}_{xy}$ are determined only by the ratio $\omega/\omega_c$ and are 
similar to MIRO. The magnetoocillations of the polarization-dependent term are more 
complicated, because they also have the magnetophonon constituent due to the
factors $\Gamma_{c2}-\tilde{\Gamma}_{c2}$ and $\Gamma_{s2}-\tilde{\Gamma}_{s2}$, 	
[see Eq. (47), Fig. 6 and its discussion below]. Therefore, the two terms in Eq. (46) 
can be distinguished from each other not only by polarization dependence and 
$B$-inversion symmetry but also by the behavior of magnetooscillations.  

It is important to emphasize that the components of the thermopower tensor 
given by Eqs. (43) and (46) do not violate the Onsager symmetry. This fact 
requires an explanation in view of the observation (see the end of Sec. III) that 
some terms in $\hat{\beta}$ violate this symmetry. Indeed, $\hat{\alpha}$ is 
formed as a result of matrix multiplication of $\hat{\rho}$ and $\hat{\beta}$ 
and its full form does contain the terms violating the Onsager symmetry. However, 
such terms are small in comparison to the terms included in Eqs. (43) and (46), 
so they are neglected.  

Coming to presentation of numerical results, let us consider first the diffusive 
contribution to thermopower coefficients. This contribution is given by Eqs. (41), (42), 
(43), and (46), where all $\Gamma_{i}$ and $\tilde{\Gamma}_{i}$ are set to zero. 
The inelastic scattering time here and below is estimated according to$^9$ 
$\tau_{in}=\varepsilon_F/T^2$. The diffusive thermopower is not sensitive to MW 
polarization. The longitudinal diffusive thermopower $\alpha_{xx}$ is modified by the 
microwaves in two ways: through the heating of 2D electrons and through the quantum 
correction in Eq. (43). The calculations (see Fig. 2) demonstrate that the heating 
mechanism is more essential. In particular, it leads to a peak at cyclotron absorption 
frequency and to oscillations at small $B$ caused by the oscillations of absorbed 
MW power due to Landau quantization. The transverse diffusive thermopower $\alpha_{xy}$, 
in contrast, is considerably affected by the MW-induced quantum corrections from 
Eq. (48). Among these corrections there is a term $\rho^{(MW)}_{xx} \beta^{(0)}_{xy}$,
whose oscillations directly reproduce the MIRO pattern shown in the inset of Fig. 2. 
The calculations demonstrate that the other terms, those in the last line of Eq. (48),
are equally important, although their contribution becomes weaker with increasing 
temperature.

\begin{figure}[ht]
\includegraphics[width=9.2cm]{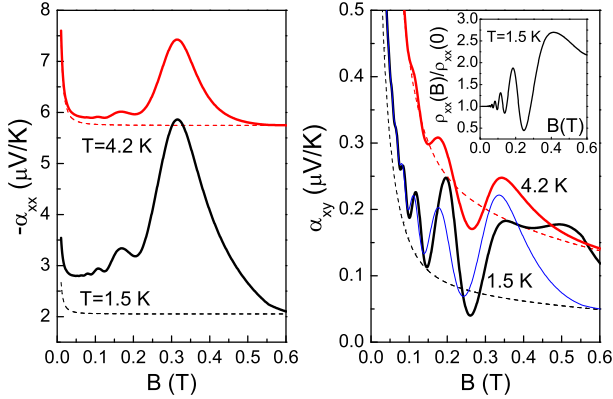}
\caption{(Color online) Longitudinal (left) and transverse (right) {\em diffusive} 
thermopower at $T=1.5$ K and $T=4.2$ K under the linearly polarized MW excitation 
of frequency 130 GHz and electric field $E_{\omega}=2$ V/cm. 
The parameters of the system are the same as in Fig. 1. The dashed lines 
show the dark thermopower (no MW excitation). The narrow solid line in the right-hand 
part shows the result of approximation $\alpha^{(MW)}_{xy} \simeq \rho^{(MW)}_{xx} 
\beta^{(0)}_{xy}$ for $T=1.5$ K. The inset presents the calculated behavior of 
the longitudinal resistance.}  
\end{figure}

Consider now the influence of microwaves on the thermopower coefficients in the 
presence of both diffusive and phonon drag mechanisms. Theoretical and experimental 
studies of GaAs quantum wells show that for the temperatures above 0.5 K the 
phonon-drag contribution dominates over the diffusive one. Consequently, the 
behavior of thermopower is governed mostly by the influence of MW excitation 
on the phonon-drag contribution. For the typical parameters of MW excitation, the 
oscillating quantum corrections given by Eq. (43) are of the order of several 
$\mu$V/K. The partial contributions due to inelastic mechanism (the first term 
in Eq. (43)) and displacement mechanism (the remaining terms) are shown in Fig. 3. 
The role of the displacement mechanism increases with increasing temperature. 
At low temperatures (Bloch-Gruneisen regime), the period of the oscillations is 
determined by the ratio $\omega/\omega_c$. With increasing temperature, the 
magnetophonon resonances become important and the picture of oscillations becomes more 
rich. The sensitivity of the displacement mechanism to MW polarization is illustrated 
by plotting its contribution for two angles of electric field of the incident wave, 
$\chi=0$ and $\chi=\pi/4$. 
     
\begin{figure}[ht]
\includegraphics[width=9.2cm]{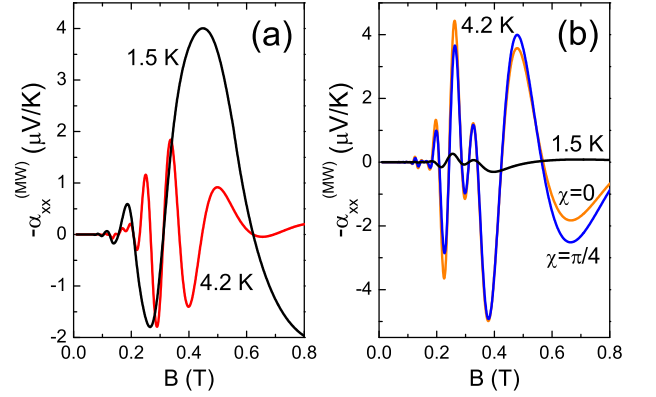}
\caption{(Color online) Microwave-induced corrections to longitudinal 
thermopower at $T=1.5$ K and $T=4.2$ K due to inelastic (a) and displacement 
(b) mechainsms, for linearly polarized MW excitation of frequency 130 GHz and 
electric field $E_{\omega}=2$ V/cm. The parameters of the system 
are the same as in Fig. 1. Two plots for $T=4.2$ K in (b) correspond 
to two angles of MW polarization.}  
\end{figure}

However, the realtive change of the longitudinal component $\alpha_{xx}$ under 
MW irradiation is not strong. The terms due to phonon drag in Eq. (43) are 
proportional to the functions $\Gamma_{s1}$, $\Gamma_{c2}$, and $\Gamma_{s2}$, 
which are small in comparison to $\Gamma_{1}$ in the important region of parameters 
$|\omega_c| < 2 p_F s_{\lambda {\bf Q}}$ and $|\omega_c| \ll 2 \pi^2 T_e$, where 
magnetophonon oscillations take place but Shubnikov-de Haas oscillations are 
suppressed (see a more detailed comparison in the Appendix). The ratio of 
the relative change of $\alpha_{xx}$ due to MW irradiation to the relative change 
of the resistivity $\rho_{xx}$ is estimated by a small factor $\Gamma_{s1}/\Gamma_{1}$. 
This means that even in the case when MW-induced resistance oscillations are 
strong, the MW-induced oscillations of the longitudinal thermopower still may 
be weak. The magnetic-field dependence of $\alpha_{xx}$ at low temperature is 
presented in Fig. 4 (a). For $T=1.5$ K one can see changes in the oscillation 
picture, in particular, inversion of the minimum around 0.18 T and a considerable 
enhancement of the last peak. The vertical shift of $\alpha_{xx}$ as a whole 
with respect to $\alpha_{xx}^{(0)}$ is caused by the diffusive mechanism 
contribution, due to heating of electrons by microwaves, see Fig. 2. With 
increasing temperature, the relative effect of microwaves on $\alpha_{xx}$ 
becomes weaker because $\alpha^{(0)}_{xx}$ increases faster than 
$\alpha^{(MW)}_{xx}$.   

\begin{figure}[ht]
\includegraphics[width=9.2cm]{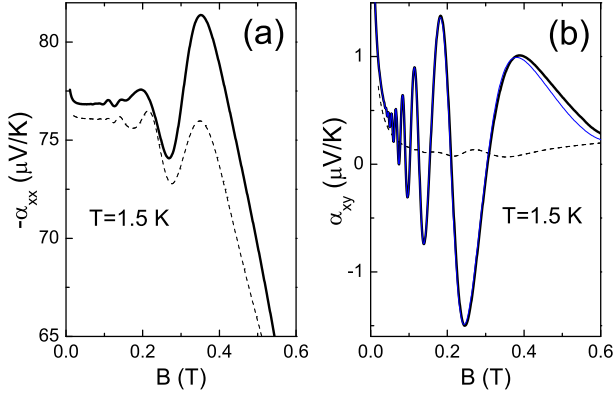}
\caption{(Color online) Longitudinal (a) and transverse (b) 
thermopower at $T=1.5$ K under the linearly polarized MW excitation 
of frequency 130 GHz and electric field $E_{\omega}=2$ V/cm. 
The parameters of the system are the same as in Fig. 1. The dashed 
lines show the dark thermopower. The narrow solid line shows the 
result of approximation $\alpha^{(MW)}_{xy} \simeq \rho^{(MW)}_{xx} 
\beta^{(0)}_{xy}$ for transverse thermopower.}  
\end{figure}

The transverse thermopower $\alpha_{xy}$, in contrast, is strongly changed 
by microwaves, because the dark thermopower $\alpha^{(0)}_{xy}$ is small 
itself. At low temperature [see Fig. 4 (b)] the modification is almost
entirely governed by the oscillations of resistivity, which means that the 
approximation $\alpha^{(MW)}_{xy} \simeq \rho^{(MW)}_{xx} \beta^{(0)}_{xy}$ 
works well. This approximation is no longer valid when temperature increases and 
the polarization-dependent contribution, the first term in the expression 
Eq. (46), becomes significant. This is demonstrated in Fig. 5, where $\alpha_{xy}$ 
is plotted for two directions of ac electric field: along $x$ axis ($\chi=0$) 
and at the angle of $\pi/4$ to this axis. With increasing $B$, when the ratio
$\nu_{tr}/\omega_c$ becomes smaller, $\alpha_{xy}$ deviates from the simple 
dependence $\propto \rho^{(MW)}_{xx}$ and becomes strongly sensitive to 
polarization. 

\begin{figure}[ht]
\includegraphics[width=9.2cm]{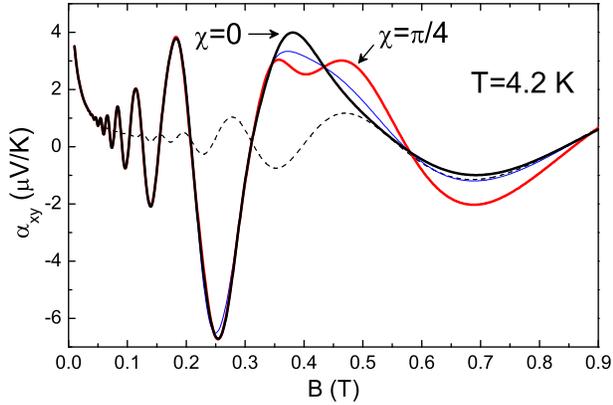}
\caption{(Color online) Transverse thermopower at $T=4.2$ K under 
the MW excitation of frequency 130 GHz and electric field 
$E_{\omega}=2$ V/cm, for two different directions of linear 
polarization of incident wave. The parameters of the system are the same 
as in Fig. 1. The dashed line shows the dark thermopower. The narrow 
solid line shows the result of approximation $\alpha^{(MW)}_{xy} \simeq 
\rho^{(MW)}_{xx} \beta^{(0)}_{xy}$.}  
\end{figure}

The polarization dependence of $\alpha_{xy}$ for different magnetic fields 
is characterized by the amplitude $\Delta \alpha_{xy}$ given by Eq. (47). This 
function is plotted in Fig. 6 for different temperatures. The complicated oscillating 
behavior of $\Delta \alpha_{xy}$ is caused by the interference of magnetophonon 
oscillations with microwave-induced oscillations. At small $T$, when the system 
is in the Bloch-Gruneisen regime, $\Delta \alpha_{xy}$ is small. With increasing 
$T$, $\Delta \alpha_{xy}$ increases and saturates around 10-15 K. The inset in 
Fig. 6 shows how the rotation of the MW polarization angle changes the total 
transverse thermopower. 

The relative contribution of polarization-dependent part can be further enhanced 
at higher MW intensity and at higher mobility, because the second term in Eq. (46)  
is proportional to the factor $\nu^2_{tr} \tau_{in}$ which goes down when inelastic 
scattering time $\tau_{in} \propto T_e^{-2}$ decreases because of microwave heating 
of electron gas and when the transport scattering rate $\nu_{tr}$ (inversely 
proportional to the mobility) decreases.

\begin{figure}[ht]
\includegraphics[width=9.2cm]{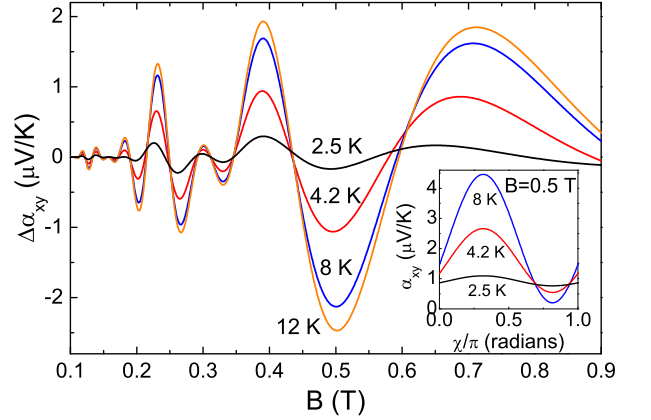}
\caption{(Color online) Magnetic-field dependence of polarization-sensitive 
part of transverse thermopower at different temperatures, for the MW excitation 
of frequency 130 GHz and electric field $E_{\omega}=2$ V/cm. 
The parameters of the system are the same as in Fig. 1. The inset shows 
dependence of thermopower on the polarization angle at $B=0.5$ T.}  
\end{figure}

In the case of circular polarization or non-polarized radiation (chaotic 
polarization) the polarization-dependent term vanishes and $\alpha^{(MW)}_{xy}$ 
is determined by the second term in Eq. (46). Since the most important part of 
this term is given by $\rho^{(MW)}_{xx} \beta^{(0)}_{xy}$, the oscillations of 
transverse thermopower under these conditions follow the MW-induced resistance 
oscillations. 
    
The longitudinal and transverse thermopower components $\alpha_{xx}$ and $\alpha_{xy}$ 
are directly measured in the Hall bars. The longitudinal thermopower can also be 
measured in the Corbino disc geometry.$^{38}$ In this case, polarization-dependent 
terms do not appear and the voltage between inner and outer contacts is determined by 
the thermopower $\alpha_{d} =\beta_{d}/\sigma_{d}$, where $\beta_{d}$ and $\sigma_{d}$ 
are the diagonal parts of the tensors $\hat{\beta}$ and $\hat{\sigma}$ in the absence 
of MW polarization. Since $\sigma_{d}$ is modified by microwaves stronger than 
$\beta_{d}$, the behavior of thermopower in MW-irradiated Corbino discs is determined 
mostly by MW-induced oscillations of $\sigma_{d}$.

The theory developed in this paper does not take into account temperature dependence 
of the density of states. Such a dependence appears mostly due to contribution of 
electron-electron scattering into the inverse quantum lifetime $1/\tau$ (see Ref. 11 
and references therein). This effect leads to an exponential suppression of all 
quantum contributions in the transport coefficients, including those considered 
above, with increasing $T_e$. Formally, this occurs because the Dingle factor $d$ 
acquires a multiplier $\exp(-\pi/\tau_{ee}(T_e)|\omega_c|)$, where $1/\tau_{ee}(T_e) 
\sim T_e^2/\varepsilon_F$. This effect tends to decrease the quantum part of dark 
thermopower and MW-induced corrections to thermopower with increasing temperature. 
Since the main (phonon-drag) contribution to thermopower, in contrast, increases with 
increasing temperature at $T < p_F s_{\lambda}$, it is important to investigate possible 
competition of these opposite trends in the quantum (proportional to $d^2$) terms in 
thermopower. Assuming that $T_e \simeq T$, the exponential dependence of these terms 
on temperature in the Bloch-Gruneisen regime ($T \ll p_F s_{\lambda}$) is written 
as $e^{-\Phi_T}$, where $\Phi_T \simeq 2 \pi T^2/\varepsilon_F |\omega_c|+ 2 p_F 
s_{\lambda}/T$ is a non-monotonic function of temperature. This function decreases 
at $T<T_0$ and increases at $T>T_0$, where $T_0 \simeq p_F s_{\lambda} (|\omega_c|/4 \pi m 
s^2_{\lambda})^{1/3}$. Since the estimate for GaAs gives $T_0 > p_F s_{\lambda}$ 
even for magnetic fields as small as 0.05 T, one may conclude that the temperature
dependence of the density of states does not alter the thermal increase of the 
quantum contributions to thermopower at $T < p_F s_{\lambda}$. However, at $T > p_F 
s_{\lambda}$ all these contributions, both in the dark thermopower and MW-induced 
corrections, decrease with temperature instead of going to saturation. 
  
\section{Discussion and conclusions}

The influence of MW irradiation on the energy distribution of electrons and on electron 
scattering by phonons and impurities has a profound effect on transport properties of 2D 
electron systems in perpendicular magnetic field. While the effect of microwaves on the 
electrical resistance is widely studied, the related behavior of the other kinetic 
coefficients has not received proper attention. This paper reports a theoretical study 
of possible MW-induced quantum effects in thermopower. Such effects can exist in the samples 
with high electron mobility in the moderately strong magnetic fields, that is, under the 
same conditions when the MW-induced quantum oscillations of the electrical resistance 
are observed. 

In contrast to electrical resistance, which at low temperatures is determined by 
electron-impurity scattering, the thermopower is determined mostly by electron-phonon 
scattering, through the phonon drag mechanism. The theory of phonon-drag thermoelectric 
response in quantizing magnetic fields remains an issue of interest even under 
quasi-equilibrium conditions, in the absence of MW irradiation. A further development 
of such theory is presented in this paper. In particular, an anisotropy of the 
acoustic phonon spectrum has been taken into account and analytical expressions 
valid in the regime of overlapping Landau levels with the accuracy up to the square 
of the Dingle factor have been derived, see Eqs. (32), (33), (41) and (42). The theory
gives a clear picture of the origin of magnetophonon oscillations observed$^{20}$ in 
the longitudinal thermopower of high-mobility GaAs quantum wells and predicts similar 
oscillations in the transverse thermopower (Fig. 1). For typical parameters of GaAs 
wells, the oscillations are clearly visible for temperatures above 2 K, while at 
lower temperatures they become exponentially suppressed because the Bloch-Gruneisen 
regime is reached. In the experiment,$^{20}$ however, the oscillations were resolved 
between 0.5 K and 1 K. This discrepancy can be explained by taking into account that 
the phonon distribution function in the experiments on thermopower is not reduced to 
the form of Eq. (17) commonly applied by theorists. Even at low temperatures of the 
sample, there can exist high-energy phonons able to cause backscattering of electrons. 
Indeed, since the phonon mean free path at low temperatures is very large (of 1 mm scale), 
it is quite possible that such high-energy phonons may arrive to the 2D system directly 
from the heater, via ballistic propagation. Another possible reason, which is especially 
relevant at low temperatures, is that the modification of phonon distribution function 
is strong and cannot be represented in the form of a small correction linear in 
temperature gradient. In any case, a quantitative agreement with experiment can 
be reached only if the phonon distribution is known. The theory presented in 
this paper can be generalized to the case of arbitrary phonon distribution by 
substituting the antisymmetric part of actual phonon distribution function 
instead of the second term in Eq. (17). 

The influence of MW irradiation on the longitudinal $\alpha_{xx}$ and transverse 
$\alpha_{xy}$ components of the thermopower has been studied above by using the 
approved methods applied earlier to calculation of the resistivity. It is found 
that the MW irradiation has a considerable effect on both these components. In 
contrast, for electrical resistance the microwaves strongly modify only the 
longitudinal component $\rho_{xx}$. Both the diffusive and 
phonon-drag contributions to thermopower are shown to be affected 
by MW irradiation. The MW-induced quantum corrections to diffusive thermopower 
increase with decreasing electron temperature, in contrast to classical 
diffusive thermopower, which is proportional to this temperature. However, 
since the phonon-drag contribution dominates, the MW-induced quantum corrections 
to phonon-drag thermopower appear to be more important. These 
effects are of the order of several $\mu$V/K for typical parameters of the 
2D system and MW excitation, and can be detected experimentally. The oscillating 
behavior of MW-induced corrections as functions of the magnetic field reflects 
the properties of electron scattering by phonons under conditions when the 
electron distribution function acquires MW-induced oscillating component 
(inelastic mechanism) and when MW-assisted scattering takes place (displacement 
mechanism). Both these mechanisms are important, and both provide a mixing 
of resonant phonon frequencies with MW frequency $\omega$, thereby leading to 
interference oscillations of the thermopower. 

In terms of relative values, the MW-induced changes in the longitudinal thermopower
are much smaller than the corresponding effect in the resistivity. In contrast, 
the relative MW-induced changes in the transverse thermopower are large, because
in the classically strong magnetic fields the transverse thermopower itself is much 
smaller than the longitudinal one. At lower temperatures and weaker magnetic fields, 
the oscillations of transverse thermopower $\alpha_{xy}$ follow the picture of 
MW-induced resistance oscillations (MIRO) [Fig. 4 (b)]. As the temperature and 
magnetic field increase, the oscillations of $\alpha_{xy}$ no longer follow the
MIRO picture and become strongly sensitive to polarization of the incident wave.
The polarization dependence of $\alpha_{xy}$ is much stronger than the corresponding 
dependence of the electrical resistivity under MW irradiation. These finding may 
stimulate experimental studies of the transverse thermopower of MW-irradiated 2D 
electron gas. 

The appearance of a large polarization-dependent term in the MW-induced 
transverse thermopower is one of the main results of the present study. The 
nature of this effect can be easily understood by considering the collisionless 
approximation (no electron-impurity scattering, $\nu_{tr}=0$), when the transverse thermopower 
does not appear without MW irradiation. The drag of electrons by the phonons drifting 
along the temperature gradient $\nabla T$ can be described$^{22}$ in terms of a dragging 
force due to effective electric field ${\bf E}_{ph} \propto \nabla T$. The 
electrons in the magnetic field are drifting perpendicular to ${\bf E}_{ph}$. To compensate 
this drift, a real electric field ${\bf E}=-{\bf E}_{ph}$ develops. 
Thus, the longitudinal thermopower is equal to $|{\bf E}|/|\nabla T|$ while 
the transverse thermopower is zero. When a polarized ac field is applied 
to the system, the effective electric field ${\bf E}_{ph}$, in general, is not directed 
along $\nabla T$ and becomes sensitive to polarization. This occurs because
${\bf E}_{ph}$ is formed as a result of electron-phonon interaction 
assisted by emission and absorption of radiation quanta, and this interaction is 
stronger when the in-plane components of phonon momenta are parallel to the 
polarization-dependent vector ${\bf R}_{\omega}$, see Eqs. (12) and (13). 
Consequently, the real electric field ${\bf E}=-{\bf E}_{ph}$ is not parallel to 
$\nabla T$, which means that there exists a transverse component of thermopower. 
This component is given by the first term in Eq. (46). Beyond the collisionless 
approximation, the other, polarization-independent terms in $\alpha_{xy}$ are also 
important. A larger relative contribution of polarization-dependent term is 
expected in 2D electron systems with higher mobility (smaller $\nu_{tr}$). 

An important issue left beyond the above consideration is the behavior of 
thermopower at zero longitudinal resistance. In high-mobility 2D systems, 
intensive MW irradiation leads to a remarkable phenomenon of zero resistance 
states,$^{3,4,5}$ which means that the longitudinal resistance vanishes in 
certain intervals of magnetic fields corresponding to MIRO minima at lower 
MW intensity. This effect is often explained (see Ref. 1 and references therein) 
as a result of the instability of homogeneous current flow under condition of 
negative local resistance, which leads to spontaneous formation of domains 
with different directions of the currents and Hall fields. Since the longitudinal 
resistivity formally enters the expression for thermopower and, as shown above, 
considerably affects the transverse thermpower in the presence of MW irradiation, 
the magnetic-field dependence should demonstrate the regions of nearly constant 
$\alpha_{xy}$ in the intervals of $\rho_{xx}=0$, while $\alpha_{xx}$ is not 
expected to be sensitive to zero resistance states. Of course, this 
conclusion looks somewhat naive, because the presence of domains may 
affect the behavior of measured thermopower. It is not clear, however, which 
kind of domain picture is realized under zero resistance state conditions in 
thermoelectric experiments, when there is no electric currents through the 
contacts. Future studies should shed light on this particularly interesting 
problem.\\ 

{\it Acknowledgement:} The author is grateful to G. Gusev for helpful discussions.\\

\appendix

\section{Asymptotic behavior of the functions $\Gamma_{i}$ and $\tilde{\Gamma}_{i}$}

In the approximation of isotropic phonon spectrum, the integral over the polar angle 
$\varphi_q$ in the operator $\hat{{\cal P}}_n$ can be carried out analytically, 
and Eq. (33) is reduced to the form
\begin{eqnarray}
\left( \begin{array}{c} \Gamma_{n} \\ \Gamma_{cn} \\ \Gamma_{sn} \end{array} \right)
= \frac{m^2}{\rho_{\scriptscriptstyle M} } \int_0^{\pi} \frac{d \theta}{\pi} (1-\cos \theta)^n
\int_{0}^{\infty} \frac{d q_z}{\pi} I_{q_z} \nonumber \\
\times \sum_{\lambda=l,t}  \tau_{\lambda} G_{\lambda} F \left(\frac{s_{\lambda} Q}{2T} \right) 
\left( \begin{array}{c} 1 \\ \cos \frac{2 \pi s_{\lambda} Q}{\omega_c} \\ 
\frac{\omega_c}{2 \pi s_{\lambda} Q} \sin \frac{2 \pi s_{\lambda} Q}{\omega_c}
\end{array} \right),
%a1
\end{eqnarray}
where $Q=\sqrt{q^2+q_z^2}$, $q=2p_F \sin(\theta/2)$, $G_{l}={\cal D}^2 + 
(eh_{14})^2 9 q^4 q_z^2/2Q^8$, and $G_{t}= (eh_{14})^2 (8 q^2 q_z^4+q^6)/2Q^8$.
For $\tilde{\Gamma}_{i}$ one should replace $G_{l}$ and $G_{t}$ by $\tilde{G}_l
=-(eh_{14})^2 9 q^4 q_z^2/4Q^8$ and $\tilde{G}_{t}=(eh_{14})^2 (8 q^4 q_z^2-q^6)/4Q^8$, 
respectively. The functions $G_t$ and $\tilde{G}_t$ describe interaction of electrons 
with transverse phonon modes due to piezoelectric potential mechanism, while $G_l$ and 
$\tilde{G}_l$ describe interaction with longitudinal phonon modes due to both deformation 
potential and piezoelectric potential mechanisms. Analytical expressions for the functions 
$\Gamma_{cn}$, $\Gamma_{sn}$, $\tilde{\Gamma}_{cn}$, and $\tilde{\Gamma}_{sn}$ 
calculated from Eq. (A1) are given below in some limiting cases. 
 
In the limit $\omega_c \ll 4 \pi s_{\lambda} p_F$, when $\cos (2 \pi s_{\lambda} Q/\omega_c)$ 
and $\sin (2 \pi s_{\lambda} Q/\omega_c)$ are rapidly oscillating functions of $\theta$ and 
$q_z/p_F$, the main contribution to the integrals in Eq. (A1) comes from the region of small 
$q_z$, when $I_{q_z} \simeq 1$, and from two regions of $\theta$ around $\theta=0$ (corresponding 
to forward scattering of electrons) and $\theta=\pi$ (backscattering), because these are 
the regions of most slow variation of $Q$ as a function of $\theta$ and $q_z$. Under 
the requirement $|\omega_c| \ll 2 \pi^2 T$, which is already stated as the condition 
when the Shubnikov-de Haas oscillations are suppressed, one obtains
\begin{eqnarray}
\Gamma_{c1}=\frac{\gamma_t}{2\epsilon_t}F\left(\frac{s_tp_F}{T}\right)\cos \epsilon_t 
-\frac{59 \gamma_t}{2^8 \epsilon_t^2} - \frac{45 \gamma_l}{2^8 \epsilon_l^2} \nonumber \\
+ 4 \frac{\gamma_l}{\epsilon_l} \left[ F\left(\frac{s_lp_F}{T}\right) \cos \epsilon_l 
+\frac{3}{\epsilon_l^3} \right] \left( \frac{{\cal D} p_F}{eh_{14}} \right)^2,
%2
\end{eqnarray}
\begin{eqnarray}
\Gamma_{c2}=\frac{\gamma_t}{\epsilon_t}F\left(\frac{s_tp_F}{T}\right)\cos \epsilon_t 
+\frac{261 \gamma_t}{2^7 \epsilon_t^4} + \frac{189 \gamma_l}{2^7 \epsilon_l^4} \nonumber \\
+ 8 \frac{\gamma_l}{\epsilon_l} \left[ F\left(\frac{s_lp_F}{T}\right) \cos \epsilon_l 
-\frac{45}{\epsilon_l^5} \right] \left( \frac{{\cal D} p_F}{eh_{14}} \right)^2,
%3
\end{eqnarray}
\begin{eqnarray}
\Gamma_{s1}=\frac{\gamma_t}{2\epsilon_t^2}F\left(\frac{s_tp_F}{T}\right)\sin \epsilon_t 
+\frac{59 \gamma_t}{2^8 \epsilon_t^2} + \frac{45 \gamma_l}{2^8 \epsilon_l^2} \nonumber \\
+ 4 \frac{\gamma_l}{\epsilon_l^2} \left[ F\left(\frac{s_lp_F}{T}\right) \sin \epsilon_l 
-\frac{1}{\epsilon_l^2} \right] \left( \frac{{\cal D} p_F}{eh_{14}} \right)^2,
%4
\end{eqnarray}
\begin{eqnarray}
\Gamma_{s2}=\frac{\gamma_t}{\epsilon_t^2}F\left(\frac{s_tp_F}{T}\right)\sin \epsilon_t 
-\frac{87 \gamma_t}{2^7 \epsilon_t^4} - \frac{63 \gamma_l}{2^7 \epsilon_l^4} \nonumber \\
+ 8 \frac{\gamma_l}{\epsilon_l^2} \left[ F\left(\frac{s_lp_F}{T}\right) \sin \epsilon_l 
+\frac{9}{\epsilon_l^4} \right] \left( \frac{{\cal D} p_F}{eh_{14}} \right)^2,
%5
\end{eqnarray}
\begin{eqnarray}
\tilde{\Gamma}_{c1}=-\frac{\gamma_t}{4\epsilon_t}F\left(\frac{s_tp_F}{T}\right)\cos \epsilon_t 
-\frac{5}{2^9}\left( \frac{\gamma_t}{\epsilon_t^2} + \frac{9 \gamma_l}{\epsilon_l^2} \right),
%6
\end{eqnarray}
\begin{eqnarray}
\tilde{\Gamma}_{c2}=-\frac{\gamma_t}{2\epsilon_t}F\left(\frac{s_tp_F}{T}\right)\cos \epsilon_t 
-\frac{21}{2^8}\left(  \frac{\gamma_t}{\epsilon_t^4} + \frac{9 \gamma_l}{\epsilon_l^4} \right),
%7
\end{eqnarray}
\begin{eqnarray}
\tilde{\Gamma}_{s1}=-\frac{\gamma_t}{4\epsilon_t^2}F\left(\frac{s_tp_F}{T}\right)\sin \epsilon_t 
+\frac{5}{2^9}\left( \frac{\gamma_t}{\epsilon_t^2} + \frac{9 \gamma_l}{\epsilon_l^2} \right),
%8
\end{eqnarray}
\begin{eqnarray}
\tilde{\Gamma}_{s2}=-\frac{\gamma_t}{2\epsilon_t^2}F\left(\frac{s_tp_F}{T}\right)\sin \epsilon_t 
+\frac{7}{2^8}\left(  \frac{\gamma_t}{\epsilon_t^4} + \frac{9 \gamma_l}{\epsilon_l^4} \right),
%9
\end{eqnarray}
where 
\begin{eqnarray}
\gamma_{\lambda}=\frac{\tau_{\lambda} m^2 (eh_{14})^2}{\pi \rho_{\scriptscriptstyle M}
p_F},~~\epsilon_{\lambda}=\frac{4 \pi s_{\lambda} p_F}{|\omega_c|}.
%10
\end{eqnarray}

For comparison, it is useful to present also the expression for $\Gamma_{1}$:
\begin{eqnarray}
\Gamma_{1}=\frac{177 \zeta(3)}{2^9} \gamma_t \left(\frac{T}{s_t p_F} \right)^2 + 
\frac{135 \zeta(3)}{2^9} \gamma_l \left(\frac{T}{s_l p_F} \right)^2 \nonumber \\
+ \left( \frac{{\cal D} p_F}{eh_{14}} \right)^2 15 \zeta(5) \gamma_l \left(\frac{T}{s_l p_F} \right)^4,~~
%11
\end{eqnarray}
where $\zeta(k)$ is the Riemann zeta-function. This expression is valid in the limit of 
$T \ll s_{\lambda} p_F$ and can be used for order-of-value estimates at $T \simeq s_{\lambda} p_F$. 

From the definition (A10), the applicability region for Eqs. (A2) - (A9) can be written as 
$\epsilon_{\lambda} \gg 1$. The magnetooscillations of the functions described by Eqs. (A2) - (A9) 
occur because of the terms with $\cos \epsilon_{\lambda}$ and $\sin \epsilon_{\lambda}$. 
The amplitudes of these oscillating terms are always much smaller than $\Gamma_{1}$ of Eq. (A11) 
in the case $\epsilon_{\lambda} \gg 1$. If $T \simeq s_{\lambda} p_F$, this smallness is given by
the factors $\epsilon^{-1}_{\lambda}$ for $\Gamma_{c1}$, $\Gamma_{c2}$, $\tilde{\Gamma}_{c1}$, 
and $\tilde{\Gamma}_{c2}$ and $\epsilon^{-2}_{\lambda}$ for $\Gamma_{s1}$, $\Gamma_{s2}$, 
$\tilde{\Gamma}_{s1}$, and $\tilde{\Gamma}_{s2}$. With lowering $T$, the oscillations are 
exponentially suppressed because of $F(s_{\lambda} p_F/T) \simeq (2s_{\lambda} p_F/T)^2 
\exp(-2 s_{\lambda} p_F/T)$ at $T \ll s_{\lambda} p_F$. In the case of strong exponential 
suppression, the absolute values of the functions given by Eqs. (A2) - (A9) are determined by 
their non-oscillating parts which are proportional to powers of $\omega_c$. The non-oscillating 
parts of $n=1$ functions ($\Gamma_{c1}$, $\Gamma_{s1}$, $\tilde{\Gamma}_{c1}$, and $\tilde{\Gamma}_{s1}$) 
are much smaller than $\Gamma_{1}$ due to parameters $(\omega_c/2 \pi^2 T)^2$ for piezoelectric-potential 
contribution and $(\omega_c/2 \pi^2 T)^4$ for deformation-potential contribution. The non-oscillating 
parts of $n=2$ functions ($\Gamma_{c2}$, $\Gamma_{s2}$, $\tilde{\Gamma}_{c2}$, and $\tilde{\Gamma}_{s2}$) 
contain extra small factors $\epsilon^{-2}_{\lambda}$, because these functions are much smaller than 
$n=1$ functions at small-angle scattering, $\theta \ll 1$.

In stronger magnetic fields, when $\omega_c$ is comparable to $4 \pi s_{\lambda} p_F$, analytical expressions
can be obtained at $T > s_{\lambda} p_F$ and under a wide-well approximation, the latter means
that the quantum well width $a$ is much larger than $\pi/p_F$ so that the convergence of the integral 
over $q_z$ takes place at $q_z \ll p_F$ and is governed by the function $I_{q_z}$. Introducing 
$q_0=\pi^{-1} \int_0^{\infty} d q_z I_{q_z}$ (for infinitely deep rectangular well $q_0=3/2a$), 
one obtains
\begin{eqnarray}
\Gamma_{cn}= \frac{2^n m^2 q_0}{\rho_{\scriptscriptstyle M}} 
\left[ (-1)^n \tau_l {\cal D}^2 {\cal I}_{2n}(\epsilon_l) \right. \nonumber  \\
+ \left. (-1)^{n-1} \tau_t \frac{(eh_{14})^2}{8p_F^2} {\cal I}_{2n-2}(\epsilon_t)  \right], 
%12
\end{eqnarray}
\begin{eqnarray}
\Gamma_{sn}= \frac{2^n m^2 q_0}{\rho_{\scriptscriptstyle M}} 
\left[ (-1)^n \tau_l \frac{{\cal D}^2}{\epsilon_l} {\cal I}_{2n-1}(\epsilon_l)
\right. \nonumber  \\
+ \left. (-1)^{n-1} \tau_t \frac{(eh_{14})^2}{8p_F^2 \epsilon_t} {\cal I}_{2n-3}(\epsilon_t)  \right], 
%13
\end{eqnarray}
\begin{eqnarray}
\tilde{\Gamma}_{cn}= - \frac{2^n m^2 q_0 (eh_{14})^2 \tau_t}{16 p_F^2 \rho_{\scriptscriptstyle M}} 
(-1)^{n-1} {\cal I}_{2n-2}(\epsilon_t), 
%14
\end{eqnarray}
\begin{eqnarray}
\tilde{\Gamma}_{sn}= - \frac{2^n m^2 q_0 (eh_{14})^2 \tau_t}{16 p_F^2 \rho_{\scriptscriptstyle M} \epsilon_t} 
(-1)^{n-1} {\cal I}_{2n-3}(\epsilon_t), 
%15
\end{eqnarray}
where 
\begin{eqnarray}
{\cal I}_{k}(x)= \frac{d^k J_0(x)}{d x^k}
%16
\end{eqnarray}
is the $k$-th order derivative of the Bessel function $J_0(x)$. Such derivatives can be 
expressed through the other Bessel functions $J_i(x)$. In the special case of $\Gamma_{s1}$, 
there is a term with the function ${\cal I}_{-1}(\epsilon_t)$, which should be treated as the
antiderivative of $J_0(\epsilon_t)$. This term is expressed through the Bessel functions 
and Struve functions ${\bf H}_i$:
\begin{eqnarray}
{\cal I}_{-1}(x) \equiv \int_0^x d x' J_0(x') = xJ_0(x) \nonumber \\
+ \frac{\pi}{2} x [J_1(x){\bf H}_0(x)-J_0(x){\bf H}_1(x)]. 
%17
\end{eqnarray}
In the regime of validity of Eqs. (A12)-(A15) the function 
$\Gamma_{1}$ is given by
\begin{eqnarray}
\Gamma_{1}= \frac{m^2 q_0}{\rho_{\scriptscriptstyle M}} 
\left[ \tau_l {\cal D}^2 + \tau_t \frac{(eh_{14})^2}{4p_F^2} \right]. 
%A18
\end{eqnarray}

For large arguments $\epsilon_{\lambda}$, the functions (A12)-(A15) are 
reduced to combinations of oscillating factors $\sin \epsilon_{\lambda}$ and 
$\cos \epsilon_{\lambda}$, similar to the case described by Eqs. (A2)-(A9),
and are small in comparison to $\Gamma_{1}$. If $\epsilon_{\lambda} \simeq 1$, 
these functions become comparable to $\Gamma_{1}$. Actually, the wide well limit 
$a \gg \pi/p_F$ is hardly attainable for single-subband occupation in the quantum 
well. The expressions (A12)-(A15) are nevertheless useful for estimates
of the maximal possible values of the quantities $\Gamma_{i}$ and $\tilde{\Gamma}_{i}$.

\end{document}